\documentclass{jpp}

\usepackage{graphicx}
\usepackage{natbib}
\usepackage[T1]{fontenc}
\usepackage{bm}
\usepackage[usenames, dvipsnames]{color}
\usepackage{graphicx}
\usepackage{amsmath}
\usepackage{amssymb}
\usepackage{amsfonts}
\usepackage{delarray}
\usepackage{courier}
\usepackage{txfonts}
\usepackage[colorlinks=true,citecolor=blue,linkcolor=blue]{hyperref}

\usepackage{color}

\ifCUPmtlplainloaded \else
  \checkfont{eurm10}
  \iffontfound
    \IfFileExists{upmath.sty}
      {\typeout{^^JFound AMS Euler Roman fonts on the system,
                   using the 'upmath' package.^^J}%
       \usepackage{upmath}}
      {\typeout{^^JFound AMS Euler Roman fonts on the system, but you
                   dont seem to have the}%
       \typeout{'upmath' package installed. JPP.cls can take advantage
                 of these fonts, if you use 'upmath' package.^^J}%
      }
  \else
  \fi
\fi

\ifCUPmtlplainloaded \else
  \checkfont{msam10}
  \iffontfound
    \IfFileExists{amssymb.sty}
      {\typeout{^^JFound AMS Symbol fonts on the system, using the
                'amssymb' package.^^J}%
       \usepackage{amssymb}%

      }{}
  \fi
\fi

\ifCUPmtlplainloaded \else
  \IfFileExists{amsbsy.sty}
    {\typeout{^^JFound the 'amsbsy' package on the system, using it.^^J}%
     \usepackage{amsbsy}}
    {\providecommand\boldsymbol[1]{\mbox{\boldmath $##1$}}}
\fi

\newsavebox{\astrutbox}
\sbox{\astrutbox}{\rule[-5pt]{0pt}{20pt}}

\newcommand{\Amp}{Amp\`{e}re's law}

\newcommand{\Av}{\mathbf{A}}

\newcommand{\Avo}{\mathbf{A}}
\newcommand{\Apa}{A_\|}
\newcommand{\Apat}{\tilde{A}_\|}

\newcommand{\Aptz}{\tilde{A}_\|^{(0)}}
\newcommand{\Ape}{A_\perp}
\newcommand{\Apet}{\tilde{A}_\perp}

\newcommand{\Ax}{A_\times}
\newcommand{\Axt}{\tilde{A}_\times}

\newcommand{\Bp}{B_p}
\newcommand{\Bpa}{B_\|}
\newcommand{\Bpat}{\tilde{B}_\|}
\newcommand{\Bpatz}{\tilde{B}_\|^{(0)}}

\newcommand{\Bv}{\mathbf{B}}
\newcommand{\bv}{\mathbf{b}}
\newcommand{\Bvz}{\mathbf{B}_0}

\newcommand{\Bz}{B_0}
\newcommand{\Bzl}{\underline{B}_0}
\newcommand{\Bvo}{\mathbf{B}_1}

\newcommand{\cd}{\cdot}

\newcommand{\dsp}{\Delta_S'}

\newcommand{\eiq}{e^{iQ}}

\newcommand{\epe}{\mathbf{e}_\perp}
\newcommand{\eps}{\epsilon}
\newcommand{\ex}{\mathbf{e}_\times}

\newcommand{\Evo}{\mathbf{E}_1}

\newcommand{\fz}{f_0}
\newcommand{\fo}{f_1}

\newcommand{\hti}{\tilde{h}_\ast}

\newcommand{\Jvo}{\mathbf{J}_1}
\newcommand{\Jz}{J_0}

\newcommand{\Jza}{J_0\left(\frac{\kpe\vpe}{\Om}\right)}
\newcommand{\Jo}{J_1}

\newcommand{\modJoa}{\frac{2J_1 (\kpe\vpe/\Om)}{\kpe\vpe/\Om}}

\newcommand{\kpe}{k_\perp}

\newcommand{\llangle}{\left\langle}

\newcommand{\na}{\nabla}
\newcommand{\nav}{\nabla_v}

\newcommand{\Om}{\Omega} 
\newcommand{\Omp}{\Omega_p}
\newcommand{\op}{\omega_p} 
\newcommand{\ord}{\mathcal{O}}
\newcommand{\ol}[1]{\overline{#1}} 
\newcommand{\ovpa}{\overline{v_\|}}

\newcommand{\p}{\partial}
\newcommand{\pho}{\Phi}
\newcommand{\pht}{\tilde{\Phi}}
\newcommand{\phz}{\Phi^{(0)}}
\newcommand{\phtz}{\tilde{\Phi}^{(0)}}
\newcommand{\pss}{\psi_\ast}

\newcommand{\rhoi}{\rho_i}
\newcommand{\rrangle}{\right\rangle}

\newcommand{\sgn}{\mathrm{sgn}}

\newcommand{\vi}{v_i}
\newcommand{\vint}{\int d^3v}
\newcommand{\vp}{\varphi}
\newcommand{\vv}{\mathbf{v}}
\newcommand{\vvpe}{\mathbf{v}_\perp}
\newcommand{\vpe}{v_\perp}
\newcommand{\vpa}{v_\|}

\newcommand{\ze}{\zeta}
\newcommand{\zev}{\hat{\zeta}}

\title[Electromagnetic zonal flow residuals]{Electromagnetic zonal flow
  residual responses}

\author[P. J. Catto, F. I. Parra, and I. Pusztai]%
{P\ls E\ls T\ls E\ls R\ns J.\ns C\ls A\ls T\ls T\ls O$^1$
\thanks{Email address for correspondence: catto@psfc.mit.edu},
\ns
F\ls E\ls L\ls I\ls X\ns I.\ns P\ls A\ls R\ls R\ls A$^{2,3}$,\ns
\and
I\ls S\ls T\ls V\ls \'{A}\ls N\ns P\ls U\ls S\ls Z\ls T\ls A\ls I$^{4}$
   }

\affiliation{$^1$
Plasma Science and Fusion Center, Massachusetts
  Institute of Technology, Cambridge, MA 02139, USA
\\[\affilskip]
$^2$ Rudolf Peierls Centre for Theoretical Physics,
  University of Oxford, Oxford, OX1 3NP, UK
\\[\affilskip]
$^3$ Culham Centre for Fusion Energy, Abingdon, OX14 3DB, UK
\\[\affilskip]
$^4$ Department of Physics, Chalmers University of 
Technology, 41296 Gothenburg, Sweden 
}

\pubyear{?}
\volume{?}
\pagerange{?}
\date{?; revised ?; accepted ?. - To be entered by editorial office}

\begin{document}

\maketitle

\begin{abstract}
The collisionless axisymmetric zonal flow residual calculation for a
tokamak plasma is generalized to include electromagnetic
perturbations.  We formulate and solve the complete initial value
zonal flow problem by retaining the fully self-consistent axisymmetric
spatial perturbations in the electric and magnetic fields.  Simple
expressions for the electrostatic, shear and compressional magnetic
residual responses are derived that provide a fully electromagnetic
test of the zonal flow residual in gyrokinetic codes.  Unlike the
electrostatic potential, the parallel vector potential and the
parallel magnetic field perturbations need not relax to flux functions
for all possible initial conditions.
\end{abstract}


\section{Introduction}
\label{intro}
A zonal flow is a sheared flow generated by turbulence that has small
scale structure compared to the system size in the radial direction
and is global in extent in the other directions. In a tokamak an
electrostatic zonal flow appears as a radially varying electric field
drift due to a radial electric field with rapid radial variation, but
with no toroidal variation. It helps reduce and regulate the turbulent
transport level in tokamaks through shear-enhanced decorrelation of
turbulent structures \citep{biglari1990, terry2000}. Its importance
was discovered when a discrepancy between gyrokinetic and gyrofluid
descriptions of ion temperature gradient (ITG) turbulence was observed
in the earliest nonlinear, electrostatic, $\delta f$, flux-tube,
particle-in-cell code (now called PG3EQ) \citep{dimits96}, where
$\delta f$ is the perturbation way from the Maxwellian. The key role
of zonal flow in controlling and reducing ITG turbulent transport,
especially near marginal stability, soon became apparent. Insights
into zonal flow behavior (missed in early gyrofluid codes) came from
code simulations and comparisons \citep{dimits2000}, leading to an
understanding that there was a nonlinear Dimits shift away from the
ITG linear stability threshold \citep{dimits96}.  \citet{RH} developed
an electrostatic analytic check to show that the zonal flow damps to a
non-zero residual level in a collisionless, axisymmetric plasma due to
polarization effects associated with the magnetic drifts.

The standard zonal flow residual calculations are electrostatic and
assume axisymmetry is maintained. An initial value problem is solved
to find the residual zonal flow level once any initial poloidal angle
dependence in the electrostatic potential is temporally damped away
via the geodesic acoustic mode (GAM) \citep{gam}. The initial
condition must normally be chosen to depend on poloidal angle to
generate a transiently evolving zonal flow or else it will be a
homogeneous solution to the non-transit averaged drift kinetic or
gyrokinetic equation. The initial distribution function is not allowed
to depend on gyro-phase since it is assumed that any initial transient
response associated with the fast gyro-motion has already damped to
its classical polarization level.  The residual zonal flow level has
proven to be an important electrostatic test of gyrokinetic codes in
general \citep{dimits2000} and the GS2 code in particular \citep{xiao2007gs2},
including collisional damping \citep{HR,xiao2007coll}, short
wavelength effects \citep{jenko2000, xiao2006FLR} and the effect of
shaping \citep{belli2006, xiao2006shape}.

In subsequent sections we generalize this axisymmetric electrostatic
model to its fully electromagnetic counterpart for a tokamak. Unlike
the procedure of \citet{terry2013}, which treats the effect of an
externally imposed, stationary (non-evolving) and non-axisymmetric
radial magnetic field perturbation on an equilibrium, we formulate and
solve a description retaining the fully self-consistent axisymmetric
spatial perturbations in the magnetic field.  The poloidal dependence
of the perturbed shear and compressional magnetic field perturbations
are retained as drives in the kinetic equation, quasineutrality, and
the parallel and perpendicular components of \Amp. The system of
equations are then solved to obtain the complete self-consistent
response within a Vlasov-Maxwell description. The expressions obtained
by solving this initial value problem provide 15 fully electromagnetic
tests of the zonal flow residual in gyrokinetic codes. The poloidally
dependent initial conditions for the fields and the distribution
function are chosen to satisfy quasineutrality and \Amp{} at $t=0$. We
assume any GAM behaviour due to poloidal variation has damped away so
that only the residual zonal flow levels are obtained.  Importantly,
the residual zonal flow levels must allow for poloidal variation of
the parallel vector potential and parallel magnetic field
perturbations for all initial conditions. The description is general
enough that even in the absence of any initial electrostatic
perturbation, a magnetic perturbation is able to generate a zonal flow
response.

The subsequent sections are organized as follows.  First, in
section~\ref{potentials} we specify the representations of the
perturbed and unperturbed fields. The kinetic equation is given in
section~\ref{kinetic}, then a suitable initial condition in terms of
perturbed fields and distribution function is chosen in
section~\ref{choice}.  The system is closed with Maxwell's equations
and solved in section~\ref{qnandamps}. Approximate expressions for the
zonal flow responses in the various fields are given in
section~\ref{sec:plots}, before we briefly summarize our results
in section~\ref{conclusions}.

%_________________________________________________
%-------------------------------------------------
%_________________________________________________

\section{Potentials, fields, and currents}
\label{potentials}
The standard electrostatic zonal flow residual calculation \citep{RH}
assumes axisymmetry is preserved during the time evolution of the
zonal flow. We seek to generalize this model to its fully
electromagnetic counterpart in a tokamak, assuming that the magnetic
field remains axisymmetric at all times. Unperturbed quantities are
assumed to evolve slowly compared to the zonal flow relaxation.

The total magnetic field is
\begin{equation}
\Bv = \Bvz + \Bvo,
\end{equation}
where 
\begin{equation}
\Bvz=I(\psi)\na\ze+\na\ze\times \na\psi=\Bz \bv
\label{Bzdef}
\end{equation}
is the background axisymmetric magnetic field, and $\Bvo$ is the
perturbed magnetic field. Here $I=I(\psi)$ must be a flux function to
make the unperturbed radial current density vanish, $2\pi\psi$ is the
unperturbed poloidal flux, $R\na\ze=\zev$ is the toroidal unit vector,
with $R$ the major radius, $\Bz = |\Bvz|$ is the magnitude of the
unperturbed magnetic field, and $\bv = \Bvz/\Bz$ is the unit vector in
the direction of the unperturbed magnetic field.

We start by representing $\Bvo$ in a form convenient to derive the
gyroaveraged kinetic equation,
\begin{equation}
\Bvo = \na \times \Avo,
\label{BvoAvo}
\end{equation}
where
\begin{equation}
\Avo=\Apa\bv+\Ape\epe+\Ax\ex.
\label{Avogyro}
\end{equation} 
Here $\epe=\na\psi/R\Bp$, $\ex = \bv \times \epe$, $|\na\psi|=R\Bp$
and $\Bp$ is the unperturbed poloidal magnetic field. We assume that
the characteristic length scale of $\Bvo$ perpendicular to the
background magnetic field $\Bvz$ is small compared to the
characteristic size of the device. Since we are considering zonal
components, the perpendicular gradient is mainly in the radial
direction. To describe this rapid radial variation of the perturbed
fields, we use the eikonal form
\begin{equation}
\left\{\Apa,\Ape,\Ax \right\}=\left\{\Apat,\Apet,\Axt
\right\}\exp[iS(\psi)],
\label{Aeikonal}
\end{equation} 
with $\na S=S'\na\psi=\kpe\epe$. The coefficients with
tilde are functions of time and are only allowed to be slow functions of
  $\psi$ and $\theta$, for example, varying as $\cos\theta$. Using equations
(\ref{BvoAvo}), (\ref{Avogyro}) and (\ref{Aeikonal}), we obtain
\begin{equation}
\Bvo \simeq \Bpa \bv - i \kpe \Apa \ex,
\label{Bvogyro}
\end{equation}
where $\Bpa = i \kpe \Ax$ is the parallel component of $\Bvo$. The
form for $\Bvo$ in (\ref{Bvogyro}) is the most common way to express
the perturbed magnetic field in gyrokinetic simulations. Note that the
component $\Ape$ of the vector potential never appears in the final
expression for $\Bvo$, and can be safely ignored.

The form for $\Bvo$ in (\ref{Bvogyro}) ensures that the magnetic field
$\Bvo$ is axisymmetric. We can make this more explicit by showing that
(\ref{Bvogyro}) is equivalent to the axisymmetric form
\begin{equation}
\Bvo = \delta\na\ze - \na\ze\times \na A,
\label{Bvodef}
\end{equation}
where $\delta(\psi, \theta)$, which need not be a flux function, is
the perturbation to $I(\psi)$, and $-A(\psi, \theta)$ is the
perturbation to $\psi$. To match equations (\ref{Bvogyro}) and
(\ref{Bvodef}), we must realize that $A$ has an eikonal form similar
to those in (\ref{Aeikonal}). Thus,
\begin{equation}
\na A \simeq \frac{i\kpe A}{R\Bp} \nabla \psi,
\end{equation}
and as a result, equation (\ref{Bvodef}) gives 
\begin{equation}
\Bvo \simeq \delta\na\ze - \frac{i\kpe A}{R\Bp} \na\ze\times \na\psi.
\label{Bvoaxi}
\end{equation}
Equation (\ref{Bvoaxi}) proves that $\Bvo$ is, to lowest order in
$(\kpe a)^{-1} \ll 1$, parallel to the flux surface. Then, it is easy
to obtain the relations between $\delta$, $A$, $\Bpa$ and $\Apa$. From
(\ref{Bvogyro}), we deduce that $\Bpa = \Bvo \cdot \bv$ and $\Apa = -
\Bvo \cdot \ex/ i \kpe$. Substituting into these two equations the
form for $\Bvo$ given in (\ref{Bvoaxi}), and using
\begin{equation}
\ex \cdot \na\ze = - \frac{\Bp}{R\Bz}
\label{exdotze}
\end{equation}
and
\begin{equation}
\ex \cdot (\na\ze \times \na\psi) = \frac{I \Bp}{R\Bz},
\label{exdotBp}
\end{equation}
we obtain
\begin{equation}
\Bpa = \Bvo \cdot \bv = \frac{I}{R^2 \Bz} \delta - \frac{i\kpe \Bp}{R \Bz} A
\label{BpavsdeltaA}
\end{equation}
and
\begin{equation}
\Apa = - \frac{1}{i \kpe} \Bvo \cdot \ex = \frac{\Bp}{i \kpe R \Bz}
\delta + \frac{I}{R^2 \Bz} A.
\label{ApavsdeltaA}
\end{equation}
Similarly, from (\ref{Bvoaxi}), we find $\delta = R^2 \Bvo\cdot
\na\ze$ and $A = - (R/i \kpe \Bp) \Bvo \cdot (\na\ze \times
\na\psi)$. Substituting into these equations the form for $\Bvo$ given
in (\ref{Bvogyro}), and using (\ref{exdotze}) and (\ref{exdotBp}), we
obtain
\begin{equation}
\delta = R^2 \Bvo \cdot \na\ze = \frac{I}{\Bz} \Bpa + \frac{i \kpe R
  \Bp}{\Bz} \Apa
\label{deltavsBpaApa}
\end{equation}
and
\begin{equation}
A = - \frac{R}{i\kpe \Bp} \Bvo \cdot (\na\ze \times \na\psi) = -
\frac{R \Bp}{i\kpe \Bz} \Bpa + \frac{I}{\Bz} \Apa.
\label{AvsBpaApa}
\end{equation}
Expressions (\ref{BpavsdeltaA}), (\ref{ApavsdeltaA}),
(\ref{deltavsBpaApa}) and (\ref{AvsBpaApa}) allow us to change from
the form most convenient for gyrokinetics, in (\ref{Bvogyro}), to the
axisymmetric form in (\ref{Bvoaxi}).

It is of interest to discuss the possible changes that the magnetic
field can undergo. It is possible to change the direction of the magnetic field lines
without changing the magnitude of $\Bv$ if $\Bpa = 0 = I\delta- i \kpe
R \Bp A $. To avoid changing the direction of the field lines, the
perturbation must satisfy $\Bvo\times\Bvz=0$, or $\Apa = 0 = R\Bp
\delta + i \kpe I A$. Changing the local theta dependent direction of the magnetic field
line does not necessarily imply a change in the safety factor 
$q(\psi) = (2\pi)^{-1} \int_0^{2\pi} (\Bv \cdot \na\ze/\Bv \cdot \na\theta)
d\theta$. To verify this we write the safety factor as
\begin{equation}
q = q_0 + q_1, 
\end{equation}
where
\begin{equation}
q_0 = \frac{1}{2\pi} \int_0^{2\pi} \frac{\Bvz \cdot \na\ze}{\Bvz \cdot
  \na\theta} d\theta
\end{equation}
is the unperturbed safety factor and
\begin{equation}
q_1 = \frac{1}{2\pi} \int_0^{2\pi} \left ( \frac{\Bvo \cdot
  \na\ze}{\Bvz \cdot \na\ze} - \frac{\Bvo \cdot \na\theta}{\Bvz \cdot
  \na\theta} \right ) \frac{\Bvz \cdot \na\ze}{\Bvz \cdot \na\theta}
d\theta.
\end{equation}
is the result of the perturbation. Note that our poloidal angle-like
variable $\theta$ is not changed by the perturbations. Using
(\ref{Bvogyro}) and (\ref{Bvoaxi}), and recalling that $\kpe/R\Bp$ is
independent of $\theta$, the perturbation $q_1$ becomes
\begin{equation}
 q_1 = \frac{1}{2\pi} \frac{i \kpe}{R\Bp} \int_0^{2\pi}
\frac{\Apa}{\bv \cdot \na\theta} d\theta = \frac{1}{2\pi} \frac{i
  \kpe}{R\Bp} \int_0^{2\pi} \left ( \frac{\Bp \delta}{i \kpe R } +
\frac{I A}{R^2 } \right ) \frac{d\theta}{\Bvz \cdot \na\theta}.
\end{equation}
When $q_1$ changes, the lines in the flux surface change topology by
switching between rational and irrational -- the only form of
reconnection allowed for axisymmetric perturbations.

%_________________________________________________
%-------------------------------------------------
%_________________________________________________

\section{Kinetic equation and solution}
\label{kinetic}

 We need to solve the linearized gyrokinetic equation in which the
 unperturbed quantities are time independent or evolve slowly compared
 to the zonal flow relaxation. The unperturbed ion distribution
 function $\fz$ is assumed to be Maxwellian:
\begin{equation}
\fz=n\left(\frac{M}{2\pi
  T}\right)^{3/2}\exp\left(-\frac{Mv^2}{2T}\right),
\label{maxwellian}
\end{equation} 
with $n$, $M$ and $T$ the ion density, mass and temperature,
respectively. Then, as we shall consider a collisionless plasma, the
linearized distribution function $\fo$ satisfies the Vlasov equation
\begin{equation}
\dot{f}_1=\frac{\p \fo}{\p
  t}+\vv\cd\na\fo+\Om\vv\times\bv\cd\nav\fo=\frac{Ze}{T}\Evo\cd\vv
\fz,
\label{perturbedvlasov}
\end{equation} 
where $\Om=ZeB_0/M c$ with $Ze$ the ion charge and $c$ the speed of
light. We have neglected the unperturbed electric field. For the
perturbed electric field we use $\Evo=-\na\pho-c^{-1}\p\Avo/\p t$,
with $\pho$ the perturbed electrostatic potential. To remove the
adiabatic piece we let $\fo=h-(Ze\pho/T)\fz$ to obtain
\begin{equation}
\dot{h}=\frac{\p h}{\p t}+\vv\cd\na h+\Om\vv\times\bv\cd\nav
h=\frac{Ze}{T} \left[\frac{\p \pho}{\p t}-\frac{\p }{\p
    t}\left(\frac{\vv\cd\Avo}{c}\right) \right] \fz.
\label{hvlasov}
\end{equation} 
This is the form that we will use to obtain the desired gyrokinetic
equation.  

Rather than perform a conventional gyrokinetic treatment
\citep{catto1978} of (\ref{hvlasov}), we use the canonical angular
momentum $\pss$ for our radial variable when we change
variables \citep{kagan1}. Then using $\vv=\vvpe+\vpa\bv$ gives
\begin{equation}
\pss=\psi-\frac{Mc}{Ze}R\zev\cd\vv=
\psi+\frac{1}{\Om}\vvpe\times\bv\cd\na\psi-\frac{I\vpa}{\Om}.
\label{psistar}
\end{equation} 
As for the vector potential in (\ref{Aeikonal}), we describe the rapid
radial variation of the perturbed electrostatic potential $\pho$ by the
eikonal expression
\begin{equation}
\pho=\pht\exp[iS(\psi)],
\label{eikonal}
\end{equation} 
where the coefficient $\pht$ is a function of time and is only allowed
to be a slow functions of $\theta$ and $\psi$. Changing from
$\psi$, $\theta$, $\ze$ and $\vv$ variables to $\pss$, $\theta$,
$\ze$, $v$, $\mu=\vpe^2/2\Bz$, and gyro-phase $\vp$, with
$\vvpe=\vpe(\epe\cos\vp+\ex\sin\vp)$, and using $\dot{\psi}_\ast=0$ to
remove the $\pss$ derivative \citep{kagan} yields the lowest order
gyrokinetic equation
\begin{equation}
\frac{\p h}{\p t}+\vpa \bv\cd\na
h=\frac{Ze\fz}{cT}\llangle\exp[iS(\psi)]\frac{\p}{\p
  t}\left[c\pht-\vpa\Apat-\vvpe\cd\left(\epe\Apet+\ex\Axt \right)
  \right] \rrangle_\vp,
\label{gk1}
\end{equation} 
where the $\theta$ dependence of $h$ is assumed slow, drift
corrections to parallel streaming are neglected as small, and the
gyroaverage $\langle\dots\rangle_\vp=(2\pi)^{-1}\oint d\vp(\dots)$ is
performed at fixed $\pss$.

Next we write $h$ in the eikonal form
\begin{equation}
h=h(\pss,\theta,v,\mu,t)=\hti
(\pss,\theta,v,\mu,t)\exp\left[iS(\pss)\right],
\label{heikonal}
\end{equation} 
where only a weak $\pss$ dependence of $\hti$ not captured by
$S(\pss)$ is allowed, $v=|\mathbf{v}|$, and
$\mu=v_\perp^2/(2B_0)$. Then, we Taylor expand to obtain
$S(\psi)-S(\pss)=Q-L+\dots$, where $L=(\kpe\vpe/\Om)\sin\vp$,
$Q=S'I\vpa/\Om=\kpe\vpa/\Omp$, and $\Omp=Ze\Bp/Mc=\Omega\Bp/\Bz$. We
retain the order $\eps$ Shafranov shift $\Delta_S$ of the flux
surfaces by writing $R=R_0(\psi)+r(\psi)\cos \theta$ with
$R_0(\psi=0)=R_0(0)$ the location of the magnetic axis,
$R_0(\psi)=R_0(0)-\Delta_S$, and $r$ the minor radius for circular
flux surfaces, then $\epsilon=r/R_0(\psi)$ is a flux function and the
ratio of the poloidal over the toroidal magnetic field is
$B_p/B_t=(\epsilon/q)[1-\Delta_S \cos \theta+\ord(\eps^2)]$ with $I=R
B_t$. As a result, the lowest order gyrokinetic equation becomes
\begin{equation}
\frac{\p \hti}{\p t}+\vpa \bv\cd\na
\hti=\frac{Ze\fz}{cT}\llangle\exp[iQ-iL]\frac{\p}{\p
  t}\left[c\pht-\vpa\Apat-\vvpe\cd\left(\epe\Apet+\ex\Axt \right)
  \right] \rrangle_\vp.
\label{gk2}
\end{equation}
Performing the gyroaverages we obtain the desired form of the ion
gyrokinetic equation
\begin{equation}
\frac{\p \hti}{\p t}+\vpa \bv\cd\na
\hti=\frac{Ze\fz}{cT}\exp(iQ)\frac{\p}{\p
  t}\left[\Jza(c\pht-\vpa\Apat)+\modJoa \frac{\vpe^2}{2\Om} \Bpat\right],
\label{gk3}
\end{equation}
where $\Bpat = i \kpe \Axt$, $\bv\cd\na=(qR)^{-1}\p/\p\theta$, the
coefficient of the $\Apet$ term has gyroaveraged to zero, and $J_0$
and $J_1$ denote Bessel functions of the first kind.

To lowest order the streaming term dominates for a weakly collisional
plasma. The damping away of any initial $\theta$ dependence leads to
the GAM behaviour observed during the early evolution to the final residual
zonal flow steady state. We are not interested in the GAM behaviour so
we annihilate the $\theta$ derivative in (\ref{gk3}) to obtain the
transit averaged gyrokinetic equation
\begin{equation}
\frac{\p \overline{\hti}}{\p t}=\frac{Ze\fz}{cT}\frac{\p}{\p
  t}\overline{\left[\Jza(c\pht-\vpa\Apat)+\modJoa \frac{\vpe^2}{2\Om} \Bpat\right]\exp(iQ)},
\label{orbitavgk}
\end{equation}
where $\overline{\hti}$ is the $\theta$ independent long time
solution, while $\hti(t=0)$ is allowed to depend on $\theta$ so we can
relate an initial $\theta$ dependent perturbation to the final steady
state $\theta$ independent solution. The transit average of any
quantity $X$ is defined as $\overline{X}=\oint d\tau X /\oint d\tau $,
with $d\tau=d\theta /(\vpa \bv\cd\na\theta)\approx
qRd\theta/\vpa$. The transit average is over a full bounce for trapped
ions and over a complete poloidal circuit for the passing ones. The
sign of $\vpa$ changes at turning points, while $d\tau>0$, giving
$\overline{\vpa}=0$ for the trapped. For the passing
$\overline{\vpa}\rightarrow \langle 1/\vpa\rangle^{-1}$ in the large
aspect ratio limit, where \\ $\langle X\rangle=[\oint X d\theta
  /(\Bvz\cd\na \theta)] [\oint d\theta /(\Bvz\cd\na \theta)]^{-1}$
denotes a flux surface average. It is easy to show that $\langle\int
d^3v\,\fz X\rangle=\langle\int d^3v\,\fz \ol{X}\rangle$. More details
are given in Appendix~\ref{AppIntegrals}.

Equation (\ref{orbitavgk}) generalizes the usual electrostatic result
to include electromagnetic effects through $\Apa$ and $\Bpa$.  It is
important to realize that $\Phi$, $\Apa$ and $\Bpa$ are allowed to be
slow functions of $\theta$ and $\psi$. Solving (\ref{orbitavgk}) by
integrating from $t=0$ gives
\begin{align}
\overline{\hti}=\overline{\hti(t=0)}+&\frac{Ze\fz}{cT}\overline{
  \left[\Jz(c\pht-\vpa\Apat)+\frac{2\Jo}{z} \frac{\vpe^2}{2\Om} \Bpat\right]\exp(iQ)}\nonumber
\\-&\frac{Ze\fz}{cT}\overline{\left\{\Jz
  \left[c\phtz-\vpa\Aptz\right]+\frac{2\Jo}{z} \frac{\vpe^2}{2\Om} \Bpatz\right\}\exp(iQ)},
\label{intgk}
\end{align}
where $\phtz=\pht(t=0)$, $\Aptz=\Apat(t=0)$ and $\Bpatz=\Bpat(t=0)$
are allowed to be flux functions, $\Jz\equiv\Jz(z)$ and
$\Jo\equiv\Jo(z)$, and $z=\kpe \vpe/\Omega$.  The strong poloidal
variation in (\ref{intgk}) is due to $\vpa$ and $Q\propto \vpa$, while
the poloidal variation of $B_0$ is weaker and sometimes unimportant.
The electron response $\overline{\tilde{h}_e}$ is given by an equation
similar to (\ref{intgk}), but with $Z\rightarrow -1$, $M\rightarrow m$
and $\fz\rightarrow f_{0e}$. We will omit species subscripts to
streamline notation except where they are needed to avoid confusion.

Next we form the perturbed quasineutrality equation
\begin{align}
0=\sum Ze\int d^3v\,
  f_{1},\label{qneq}
\end{align}
 and the two components
of \Amp{}
\begin{align}
\kpe^2\Apa \simeq (4\pi/c)\Jvo\cd\bv=(4\pi/c)\sum Ze\int d^3v\,
  f_{1}\vpa\label{eq32}
\end{align}
and 
\begin{align}
\kpe^2\Ax\simeq (4\pi/c)\Jvo\cd\ex=(4\pi/c)\sum Ze\int d^3v\,
  f_{1}\vpe\sin\vp\label{eq33},
\end{align}
where the integrals are taken at fixed $\psi$, $\sum$ denotes a sum
over ions and electrons. Notice that $\Jvo\cd\epe \simeq 0$ since
$\na\cd\Jvo=0=\na\cd \Av$.  Using $\fo=\hti
\exp[iS(\pss)]-(Ze\fz/T)\pht \exp[iS(\psi)]$ and $\vint h
\left\{1,\vpa,\vpe\sin\varphi\right\} =\vint \tilde{h}_\ast
\left\{1,\vpa,\vpe\sin\varphi\right\} \exp[i(L-Q+S)] =\vint
\tilde{h}_\ast \left\{\Jz,\vpa \Jz,i\vpe \Jo\right\} \exp[i(S-Q)] $,
equations (\ref{qneq}), (\ref{eq32}) and (\ref{eq33}) give
\begin{align}
\sum Ze\int d^3v\,
  \tilde{h}_\ast J_0 e^{-iQ} - \sum \frac{Z^2 e^2 n}{T} \pht = 0,\label{gkqneq}
\end{align}
\begin{align}
\kpe^2\Apat = \frac{4\pi}{c} \sum Ze\int d^3v\,
  \tilde{h}_\ast \vpa J_0 e^{- i Q} \label{gkeq32}
\end{align}
and 
\begin{align}
\frac{\Bz \Bpat}{4\pi} + \sum \int d^3v\,
  \tilde{h}_\ast \frac{M \vpe^2}{2} \frac{2J_1}{z} e^{- i Q} = 0. \label{gkeq33}
\end{align}
Note that the perpendicular \Amp{} (\ref{eq33}) has become perpendicular pressure balance (\ref{gkeq33}).

Equation (\ref{intgk}) represents the formal solution of an initial
value problem for a time scale long compared to the periodic
gyromotion about guiding centers. It depends on the initial value
$\hti(0)$, which we are free to choose arbitrarily as long as it
satisfies Maxwell's equations. Our particular choice for $\hti(0)$
will be motivated in the next section.

%_________________________________________________
%-------------------------------------------------
%_________________________________________________

\section{Choice of initial condition}
\label{choice}

 We must pick $\hti$ to satisfy the Maxwell equations at $t=0$, while
 also obtaining a convenient and sensible form for $\overline{\hti}$
 to evaluate the long time relaxation behavior of the zonal flow
 response. The non-transit averaged initial condition for $\hti(t=0)$
 must depend on $\theta$ as well as $\pss$, $v$, $\mu$ to generate a
 transiently evolving zonal flow (the GAM) or else it will be a
 homogeneous solution to the non-transit averaged Vlasov
 equation. However, we do not allow $\hti(t=0)$ to depend on gyrophase
 since we assume any initial transient response associated with
 gyromotion has already damped to its classical polarization level.

We require Maxwell's equations to be satisfied at $t=0$, which
  we take to mean after many gyrations, but much less than the time
  for a poloidal bounce or transit to be completed. Consequently, at
  $t=0$ we must satisfy (\ref{gkqneq})-(\ref{gkeq33}).

A GAM develops on a time scale of the order of a transit or bounce
time and much longer than a gyration period. It oscillates as it damps
away to the residual zonal flow level. The initial conditions are
sometimes viewed as approximating the turbulent sources \citep{RH,
  sugama} of charge and current densities on a time much less than a
transit or bounce time, but after many gyrations.  When the long time
behavior of the system is studied electrostatically we may use the
transit average result
\begin{equation}
\fo=\ol{\hti(t)}e^{i S(\pss)}-\frac{Ze\fz}{T}\pht e^{i S(\psi)}, 
\label{taapprox}
\end{equation}
where $\ol{\hti(t)}$ is given by the electrostatic limit of
(\ref{intgk}).

Next we explain how we choose $\hti(0)$ to include magnetic
perturbations.  For our results to have the required generality we do
not flux surface average quasineutrality and the two components of
\Amp. Moreover, we desire forms at $t = 0$ in which $\phtz$, $\Aptz$
and $\Bpatz$ terms only contribute to quasineutrality, and the
parallel and perpendicular components of \Amp, respectively. To avoid
the need for complicated velocity space structure, we do not consider
arbitrary wavenumbers \citep{xiao2007gs2,xiao2006FLR}.  We do
manipulations consistent with an expansion in $\kpe\vpe/\Om\ll 1$, but
to avoid lengthy expressions, we prefer to keep the finite Larmor
radius terms in the form of Bessel functions for a while and make the
expansions explicit later.  To simplify our treatment and properly
recover the $\kpe\vpe/\Omega \ll Q \ll 1$ limit electrostatically we
can use the simple form
\begin{align}
\hti(0)=\frac{Ze\fz \eiq}{T\Jz}\phtz,
\label{initialES}
\end{align}
where since we are only interested in $\kpe\vpe/\Omega \ll 1$ the
zeroes of the Bessel function are of no concern. A nice discussion of
the difference between treating electrostatic zonal flows as an
initial value problem rather than as a turbulent source in
quasineutrality is presented in Sec. 3 of \cite{monreal16}. Moreover,
their paper and the references therein should be consulted to
understand the differences between the residual zonal flow behaviour in
stellarators and tokamaks.

The preceding expression motivates us to assume
\begin{align}
\hti(0)=\frac{Ze\fz \eiq}{T\Jz}
\left[H\phtz+\frac{\kpe^2 c}{\op^2}\vpa\Aptz-\frac{
    2Tu}{Ze\Bz\beta}\Bpatz \right],
\label{initial}
\end{align}
where we have introduced the species dependent quantities
\begin{equation}
H=1-\frac{3\alpha}{2} \left(\frac{Mv^2}{3T}-1\right)
\label{Hdef}
\end{equation}
and
\begin{equation}
\alpha=\frac{2 \vint  \fz \vpe^2 \left[\frac{2\Jo}{z\Jz}-1\right]}{\vint
  \fz  \vpe^2 \frac{2\Jo}{z\Jz} \left(\frac{Mv^2}{T}-3\right)},
\label{alphadef}
\end{equation}
with $z=\kpe\vpe/\Omega$, $2\Jo/z\Jz\approx 1+z^2/8$ giving $\alpha\approx \kpe^2 T/2M\Omega^2\ll 1$ for $z\ll 1$, and the species independent quantity
\begin{equation}
u=\frac{\sum nT}{\sum \vint \fz M\vpe^2\frac{\Jo}{z\Jz} }.
\label{udef}
\end{equation}
In the preceding $\beta=8\pi\sum n T
/\Bz^2=8\pi(n_iT_i+n_eT_e)/\Bz^2=\beta_i+\beta_e$ and $\op^2=4\pi\sum
Z^2e^2 n/M=4\pi
e^2[(Z^2n_i/M)+(n_e/m)]=\omega_{pi}^2+\omega_{pe}^2$. We will also
make use of the definitions $\vi=(2T_i/M)^{1/2}$,
$\rhoi=\vi/\Omega_i$, $\Omega_i=ZeB_0/(Mc)$, and
$\rho_{pi}=q\rhoi/\eps$. 

The functional form of $H$ in
(\ref{initial}) is chosen so at $t = 0$ it does not alter
quasineutrality  (\ref{gkqneq}) since
\begin{equation}
\phtz \sum \frac {Z^2 e^2}{T}n=\phtz \sum \frac {Z^2 e^2}{T}\vint \fz
H=\phtz \sum \frac {Z^2 e^2}{T}\vint \fz.
\label{qnattz}
\end{equation}
Moreover the $\alpha$ in $H$ is chosen so $\phtz$ will not enter
perpendicular \Amp{} by taking
\begin{equation}
\phtz\sum \frac{ZeM}{2T}\vint \fz\vpe^2\left[H \frac{2\Jo}{z\Jz}-1\right]=0.
\label{alphafrom}
\end{equation}
The $\Aptz$ term does not enter quasineutrality because its integral
is odd in $\vpa$, and $\Bpatz$ does not enter because $u$ and $\beta$
are species independent as we can use unperturbed quasineutrality. We
can also see that parallel \Amp, (\ref{gkeq32}), is satisfied at $t =
0$ since the $\phtz$ and $\Bpatz$ integrals are odd in $\vpa$ leaving
\begin{equation}
\Aptz=\frac{4\pi}{\op^2}\Aptz\sum \frac{Z^2e^2}{T} \vint \fz\vpa^2
=\frac{\Aptz}{\op^2}\sum \frac{4\pi Z^2e^2n}{M}.
\label{alphafrom2}
\end{equation}
To satisfy the perpendicular \Amp, (\ref{gkeq33}), at $t = 0$
we require (\ref{Hdef}) to (\ref{udef})  to be satisfied as can be seen from 
\begin{equation}
\Bz\Bpatz=-4\pi\sum\vint \fz M\vpe^2(\Jo/z\Jz)
\left[\frac{Ze}{T}H\phtz-\frac{2u}{\Bz\beta}\Bpatz \right]
=\Bpatz\frac{u}{\beta}\frac{8\pi\sum nT}{\Bz u},
\label{bpaic}
\end{equation}
where we use unperturbed quasineutrality as well as $\vint \fz
\vpe^2[(2\Jo H/z\Jz)-1]=0 $.

Apart from the need to satisfy Maxwell's equations, $\hti(t=0)$ is
arbitrary.  Other choices for $\hti(t=0)$ may be used, but
  (\ref{initial}) is sufficient for our purpose. It has the
transit average
\begin{align}
\overline{\hti(0)}=&\frac{Ze\fz}{T}\left[H\phz\ol{\eiq/\Jz}-\frac{2
    Tu}{Ze\Bz^2\beta}\Bpatz \ol{\Bz\eiq/\Jz}+ \frac{\kpe^2c}{\op^2}
  \Aptz\ol{\vpa\eiq/\Jz}\right] ,
\label{initialaverage}
\end{align}
where the weak $\theta$ dependences of $u$ and $\alpha$ are
neglected.

Before closing this section we prove that $\pht$ is a lowest
  order flux function for fully electromagnetic initial
  conditions. For this demonstration we will only be interested in the
  $L\sim\kpe\vpe/\Omega\ll Q\sim \kpe \vpa/\Omp \ll 1$
  limit. Consequently, to lowest order we let $\kpe\rightarrow 0$ to
  make $\eiq\rightarrow 1$, $\Jz\rightarrow 1$, $\Jo\rightarrow \kpe
  \vpe/2\Om$, $\alpha\rightarrow 0$, $H\rightarrow 1$ and
  $u\rightarrow 1$ in (\ref{gkqneq})-(\ref{gkeq33}), and then use
  $\hti\approx\ol{\hti}$ to obtain quasineutrality in the form
\begin{equation}
\pht\sum\frac{Z^2e^2}{T}n=\sum \frac{Ze}{T}\vint \ol{\hti},
\label{qnlw1}
\end{equation}
with
\begin{equation}
\ol{\hti(t)}=\ol{\hti(0)}+\frac{Ze}{cT}\fz\left[   
\left(c\ol{\pht}-\ol{\vpa \Apat}+\vpe^2\ol{\Bpat}/2\Om \right)
-\left(c\phtz-\ol{\vpa}\Aptz+\Bpatz\vpe^2/2\Om \right)
\right],
\label{htlw1}
\end{equation}
and 
\begin{equation}
\ol{\hti(0)}=\frac{Ze}{T}\fz\left[   
\phtz+\frac{\kpe^2c}{\op^2}\ovpa\Aptz-\frac{2T\ol{\Bz}}{Ze\Bz^2\beta}\Bpatz
\right],
\label{htzlw1}
\end{equation}
where we have set $\alpha=0$ and $u = 1$ since we let $\kpe\rightarrow
0$ and used $\ol{\vpe^2/2\Bz}=\mu=\vpe^2/2\Bz$. The integrals odd in
$\vpa$ do not contribute to quasineutrality, of course. In addition,
unperturbed quasineutrality prevents the $\Bpatz$ and $\Bpat$ from
contributing to perturbed quasineutrality as well since they result in
the integrals $\sum Ze \vint\fz\ol{\Bz}=(\Bz/2\Bzl)(\int_0^{\Bzl/\Bz} d\lambda
\ol{\Bz}/\xi)\sum Zen=0$ and 
\begin{align}
\sum\frac{ZeM}{T}\vint\fz\mu\ol{\Bpat}= & \frac{3\Bz}{2\ol{\Bz}^2} \sum
Zen \left(\int_0^{\Bzl/\Bz} d\lambda \lambda \frac{\ol{\Bpat}}{\xi}
\right) \nonumber \\ = & \frac{3\Bz}{2\ol{\Bz}^2}\left(\int^{\Bzl/\Bz}_0 d\lambda
\lambda \frac{\ol{\Bpat}}{\xi}\right) \sum Zen=0,
\label{zeroint1}
\end{align}
where $\Bz/\Bzl\simeq R_0
(\psi)/R=1-\epsilon(\psi)\cos\theta+\ord(\epsilon^2)$,
$\Bzl=\langle\Bz\rangle$, $\epsilon=r(\psi)/R_0(\psi)$,
$\lambda=2\mu\Bzl/v^2$, $\xi=\sqrt{1-\lambda \Bz/\Bzl}$, and $d^3v\rightarrow \sum_{\sgn \vpa} \pi\Bz v^3dv
  d\lambda/\Bzl|\vpa|$ (where $\sum_{\sgn \,\vpa}$ can be replaced by
  a factor 2 when the integrand is even in $\sgn\, \vpa$).
Consequently, assuming $\kpe\vpe/\Om\ll Q\ll 1$ the lowest order
perturbed quasineutrality equation becomes
\begin{equation}
\pht\sum\frac{Z^2e^2}{T}n=\sum \frac{Ze}{T}\vint \ol{\hti}=
\sum \frac{Z^2e^2}{T}\vint \fz \ol{\pht},
\label{qnlw2}
\end{equation}
which requires
\begin{equation}
\pht=\frac{\vint\, \ol{\pht} \fz }{\vint \fz},
\label{phiff}
\end{equation}
and is satisfied if $\pht=\langle\pht\rangle$. Therefore, we may
safely assume $\pht$ is a flux function to lowest order. To see this
more rigorously, we multiply (\ref{qnlw2}) by $\pht$ to form
$\pht^2\vint \fz=\vint \fz \pht^2=\pht\vint\fz\ol{\pht}=\vint\fz\ol{\pht}\pht$,
then flux surface average to obtain
\begin{equation}
\llangle \vint \fz\pht^2 \rrangle=\llangle \vint \fz\ol{\pht}\pht
\rrangle.
\label{phiff2}
\end{equation}
As a result, using $\llangle \vint \fz\pht^2 \rrangle=\llangle \vint
\fz\ol{\pht^2} \rrangle$ and $\llangle \vint \fz\ol{\pht}\pht
\rrangle=\llangle \vint \fz\ol{\pht}^2 \rrangle$ we find
\begin{equation}
0=\llangle \vint \fz \left(\ol{\pht^2}-\ol{\pht}^2\right)\rrangle=\llangle \vint \fz \ol{\left(\ol{\pht}-\pht\right)^2}\rrangle.
\label{phiff3}
\end{equation}
Therefore, we need
\begin{equation}
\pht=\ol{\pht},
\label{phiff4}
\end{equation}
but $\pht=\pht(\psi,\theta)$, while $\ol{\pht}=\ol{\pht}(\psi,\lambda)$, so it must be that
\begin{equation}
\pht=\langle\pht\rangle
\label{phiff5}
\end{equation}
to lowest order.  

To generalize our results for quasineutrality and to treat both
components of \Amp, we extend our initial condition (\ref{htzlw1}) to
include finite orbit effects by employing (\ref{initialaverage}) in
(\ref{intgk}).

In the calculation of the electrostatic zonal flow residual \citep{RH} the
non-adiabatic electron response could be neglected as small in
$(m/M)^{1/2}$, where $m$ is the electron mass, however here it
sometimes needs to be retained, since the electromagnetic terms in the
non-adiabatic response are proportional to the thermal speed of the
species.  In the next section we will use the preceding results to
form quasineutrality and \Amp.

%_________________________________________________
%-------------------------------------------------
%_________________________________________________

\section{Quasineutrality and \Amp}
\label{qnandamps}

In this section we form and consider the non-flux surfaced average
components of \Amp{} to demonstrate that poloidal variation of
the parallel vector potential and the parallel magnetic field must be
retained. Indeed, we will find that there are cases for which these
field responses have strong poloidal variation. In addition, we will
perform a more complete evaluation of quasineutrality once we have
examined the two components of \Amp.

To perform the derivation of the two components of \Amp{} we
must realize that $\Apat$ and $\Bpat$ are not normally flux
functions. Indeed, even when $\tilde B_{||}^{(0)} = 0 = \tilde
A_{||}^{(0)}$ we will find they both have poloidal
variation. Retaining $\eiq$ modifications, but ignoring $L\sim
\kpe\vpe/\Om$ corrections as unimportant except in the $\tilde \Phi $
and ${\tilde \Phi ^{(0)}}$ terms, we use
\begin{equation}
\overline {{{\tilde h}_ \ast }(t)} = \overline {{{\tilde h}_ \ast
  }(0)} + \frac{{Ze}}{{cT}}{f_0}\left[c\overline {\tilde \Phi
    {J_0}{e^{iQ}}} + \frac{{v_ \perp ^2}}{{2\Omega }}\overline
  {{{\tilde B}_{||}}{e^{iQ}}} - \overline {{{\tilde
        A}_{||}}{v_{||}}{e^{iQ}}} - c{\tilde \Phi ^{(0)}}\overline
  {{J_0}{e^{iQ}}} - \frac{{v_ \perp ^2}}{{2\Omega }}\tilde
  B_{||}^{(0)}\overline {{e^{iQ}}}   + \tilde A_{||}^{(0)}\overline
  {{v_{||}}{e^{iQ}}} \right],
\end{equation}
with 
\begin{equation}
\overline {{{\tilde h}_ \ast }(0)} = \frac{{Ze}}{T}{f_0}\left[ {\tilde
    \Phi ^{(0)}}\overline {({e^{iQ}}H /  {J_0})} +
  \frac{{k_ \perp ^2c}}{{\omega _p^2}}\overline {{v_{||}}{e^{iQ}}}
  \tilde A_{||}^{(0)} - \frac{{2T}}{{ZeB_0^2\beta }}\tilde
  B_{||}^{(0)}\overline {{B_0}{e^{iQ}}} \right].
\label{eq5p2}
\end{equation}

\subsection{\Amp}
\label{sec:amps}
  
Inserting the preceding into  
\begin{equation}
{\tilde B_{||}} =  - \sum \frac{{2\pi M}}{{{B_0}}}\int {d^3}v\overline {{{\tilde h}_ \ast }} v_ \perp ^2{e^{ - iQ}}
\end{equation}
and
\begin{equation}
k_ \perp ^2\Apat = \frac{{4\pi }}{c}\sum Ze\int {d^3}v\overline {{{\tilde h}_ \ast }} {v_{||}}{e^{ - iQ}},
\label{eq5p4}
\end{equation} 
we obtain
\begin{align}
{\tilde B_{||}} =  - &\sum \frac{{4\pi Ze}}{{{B_0}}}\int
{d^3}v{f_0}\frac{{Mv_ \perp ^2}}{{2T}}\frac{{2{J_1}}}{{z }}{e^{ -
    iQ}}\left[\overline {\tilde \Phi {J_0}{e^{iQ}}} + {\tilde \Phi
    ^{(0)}}(\overline {{e^{iQ}}H / {J_0}} - \overline {{J_0}{e^{iQ}}}
  )\right]\nonumber 
\\ + & \sum \frac{{4\pi Ze}}{{c{B_0}}}\int
{d^3}v{f_0}\frac{{Mv_ \perp ^2}}{{2T}}{e^{ - iQ}}\left[\overline
  {{v_{||}}{{\tilde A}_{||}}{e^{iQ}}} - \tilde A_{||}^{(0)}(1 +
  \frac{{k_ \perp ^2{c^2}}}{{\omega _p^2}})\overline
       {{v_{||}}{e^{iQ}}} \right]
	\label{eq5p5} 
\\ - & \sum \frac{{4\pi T}}{{B_0^2}}\int {d^3}v{f_0}\frac{{Mv_ \perp
    ^2}}{{2T}}{e^{ - iQ}}\left[\frac{{Mv_ \perp ^2}}{{2T}}(\overline
  {{{\tilde B}_{||}}{e^{iQ}}} - \tilde B_{||}^{(0)}\overline
  {{e^{iQ}}} ) - \frac{{2  \tilde B_{||}^{(0)}}}{{{B_0}\beta
  }}\overline {{B_0}{e^{iQ}}} \right]\nonumber
\end{align}
and 
\begin{align}\Apat = -& \sum \frac{{4\pi {Z^2}{e^2}}}{{k_ \perp
      ^2{c^2}T}}\int {d^3}v{f_0}{v_{||}}{e^{ - iQ}}\left[\overline
    {{v_{||}}{{\tilde A}_{||}}{e^{iQ}}} - \tilde A_{||}^{(0)}(1 +
    \frac{{k_ \perp ^2{c^2}}}{{\omega _p^2}})\overline
         {{v_{||}}{e^{iQ}}} \right]\nonumber \\ +& \sum \frac{{4\pi
      {Z^2}{e^2}}}{{k_ \perp ^2cT}}\int {d^3}v{f_0}{v_{||}}{e^{ -
      iQ}}{J_0}\left[\overline {\tilde \Phi {J_0}{e^{iQ}}}   +  {\tilde
      \Phi ^{(0)}}(\overline {{e^{iQ}}H / {J_0}} - \overline
    {{J_0}{e^{iQ}}} )\right]\label{eq5p6} \\ +& \sum \frac{{4\pi Ze}}{{k_
      \perp ^2c{B_0}}}\int {d^3}v{f_0}{v_{||}}{e^{ - iQ}}\left[\frac{{Mv_
        \perp ^2}}{{2T}}(\overline {{{\tilde B}_{||}}{e^{iQ}}} -
    \tilde B_{||}^{(0)}\overline {{e^{iQ}}} ) - \frac{{2 \tilde
        B_{||}^{(0)}}}{{{B_0}\beta }}\overline {{B_0}{e^{iQ}}}
  \right]\nonumber.
\end{align}

Recalling that $\Apat$ and ${\tilde B_{||}}$ need not be flux
functions and that the unperturbed magnetic field gives rise to $\cos
\theta $ dependence, we Fourier decompose and retain only the leading
poloidal dependence by writing
\begin{equation}
\Apat = \llangle \tilde{A}\rrangle + a\cos \theta + \dots\simeq
\left[\llangle \Apat\rrangle -\frac{a}{2}(\eps-\Delta_S')\right]+
a\cos \theta
\label{eq5p7}
\end{equation} 
and 
\begin{equation}{\tilde B_{||}} = \llangle \tilde{B} \rrangle + b\cos \theta + \dots \simeq \left[\llangle
    {\tilde B_{||}}\rrangle -\frac{b}{2}(\eps-\Delta_S')\right]+ b\cos
  \theta
\label{eq5p8}
\end{equation}
with the coefficients $\langle \Apat\rangle$, $\langle {\tilde
  B_{||}}\rangle$, $\langle \tilde{A}\rangle$, $\langle
\tilde{B}\rangle$, $a$, and $b$ flux functions, and $\Bvz\cd\na
\theta\propto 1-(\eps-\Delta_S') \cos\theta$ in the flux surface
averages for our Shafranov shifted circular flux surface model, where
$\dsp=d\Delta_S/dr\sim \eps$ . We do not assume an ordering of $a$ and
$b$ relative to $\langle \Apat\rangle$ and $\langle {\tilde
  B_{||}}\rangle$ as we will calculate all four of these coefficients.
In addition, we recall the electrostatic result of \citet{RH},
\begin{equation}
\llangle \tilde \Phi \rrangle   = \frac{{{{\tilde \Phi
      }^{(0)}}}}{{1 + \gamma {q^2}/ {\epsilon ^{1/2}}}},
\label{rhresult}
\end{equation} 
with $\gamma\approx 1.64$. The preceding allows us to assume
${q^2}{\epsilon ^{ - 1/2}}\llangle \tilde \Phi \rrangle \sim{\tilde
  \Phi ^{(0)}}$ for $\beta\ll 1$ and $\epsilon \ll 1$.  

Expanding for $Q=\kpe\vpa/\Omp\ll 1$, inserting $H$, neglecting
$\eps$ corrections to $k_ \perp ^2\rho _i^2$ terms, ignoring all
$Q^2$, $\beta$ and $\eps^{3/2}$ corrections to the ${\tilde B_{||}}$
and $\tilde B_{||}^{(0)}$ terms in (\ref{eq5p5}), recalling $\tilde
\Phi = \llangle \tilde \Phi \rrangle $, and using unperturbed
quasineutrality, perpendicular \Amp{} becomes
\begin{align}
{\tilde B_{||}} - \tilde
B_{||}^{(0)} \frac{\Bzl}{B_0}  = & - \langle \tilde \Phi \rangle \sum
\frac{{2\pi ZeM}}{{T{B_0}}}\int {d^3}v{f_0}v_ \bot ^2\left[\overline Q
   Q - \frac{({Q^2}  + \overline {{Q^2}} ) }{2}\right] + (\langle \tilde \Phi \rangle - {\tilde \Phi
  ^{(0)}})\frac{{3c{\beta _i}k_ \bot ^2}}{{4{\Omega _i}}}\nonumber
\\ + & i{k_ \perp }\sum \frac{{2\pi q{M^2}}}{{\epsilon TB_0}}\int
   {d^3}v{f_0}v_ \perp ^2(\overline{\Bz^{-1} v_{||}^2{{\tilde A}_{||}}} -
   \Bz^{-1}{v_{||}}\overline {{v_{||}}{{\tilde A}_{||}}}  )\label{eq5p10} 
\\-& i{k_ \perp
   }\tilde A_{||}^{(0)}\left(1 + \frac{{k_ \perp ^2{c^2}}}{{\omega
       _p^2}}\right)\sum \frac{{2\pi q{M^2}}}{{\epsilon TB_0}}\int
   {d^3}v{f_0}v_ \perp ^2(\overline {\Bz^{-1}v_{||}^2} - \Bz^{-1}{v_{||}}\overline
   {{v_{||}}} ).\nonumber
\end{align} 
Here and hereafter, we use $Q\propto \vpa/\Bz$ and then treat $\kpe$
as a flux function, except in $k_\perp^2\tilde A_{||}^{(0)}$ terms
from (\ref{eq5p2}) and $k_\perp^2 \tilde A_{||}$ from (\ref{eq5p4}).

In (\ref{eq5p6}) we neglect $Q$ corrections for $\Apat$ and $\tilde
A_{||}^{(0)}$ terms, insert $H$, and continue to use $\tilde \Phi =
\llangle \tilde \Phi \rrangle $, to find that parallel \Amp{} is
\begin{align}
\Apat +& \sum \frac{{4\pi {Z^2}{e^2}}}{{k_ \perp ^2{c^2}T}}\int
      {d^3}v{f_0}{v_{||}}\overline {{v_{||}}{{\tilde A}_{||}}} -
      \tilde A_{||}^{(0)}\left(1 + \frac{{k_ \perp ^2{c^2}}}{{\omega
          _p^2}}\right)\sum \frac{{4\pi {Z^2}{e^2}}}{{k_ \perp
          ^2{c^2}T}}\int {d^3}v{f_0}{v_{||}}\overline {{v_{||}}}
      \nonumber 
\\=& i\sum \frac{{4\pi qM}}{{{k_ \perp }\epsilon
          B_0}}\int {d^3}v{f_0}\frac{{Mv_ \perp
          ^2}}{{2T}}{v_{||}}\left[(\overline {\Bz^{-1}{v_{||}}{{\tilde
              B}_{||}}}   - \Bz^{-1}{v_{||}}\overline {{{\tilde B}_{||}}} ) -
        \tilde B_{||}^{(0)}(\overline {\Bz^{-1}{v_{||}}} -
              \Bz^{-1} {v_{||}})\right]\nonumber 
\\ -& i\frac{2}{\beta }\tilde
      B_{||}^{(0)}\sum \frac{{4\pi qM}}{{{k_ \perp }\epsilon
          B_0^2}}\int {d^3}v{f_0}{v_{||}}(\overline {{v_{||}}}  
      -  {v_{||}})\nonumber 
\\ +& i\llangle \tilde
      \Phi \rrangle \sum \frac{{2\pi {Z^2}{e^2}}}{{k_ \perp ^2cT}}\int
           {d^3}v{f_0}{v_{||}}\left[Q\overline {{Q^2}} -
             {Q^2}\overline Q + \frac{{{Q^3} - \overline {{Q^3}}
             }}{3}\right] \label{eq5p11} 
\\ +& i \sum \frac{{2\pi
               c{M^2}}}{{TB_0^2}}\left[\left(\llangle \tilde \Phi
             \rrangle - {\tilde \Phi ^{(0)}}\right)\int
                      {d^3}v{f_0}{v_{||}}(Q - \overline Q )v_ \perp ^2
                      + {\tilde \Phi ^{(0)}}\frac{T}{M} \int
                      {d^3}v{f_0}{v_{||}}(Q - \overline Q
                      )\right],\nonumber
\end{align}
where we used (\ref{A22}), (\ref{A26}), and (\ref{A31}) and similar
integrals for $\llangle \tilde \Phi \rrangle $ and ${\tilde \Phi
  ^{(0)}}$ terms to see that due to unperturbed quasineutrality$\sum
({Z^2}{e^2}/ T)\int {d^3}v{f_0}{v_{||}}(Q - \overline Q ) = 0$. The
zonal flow responses of the perturbed parallel magnetic field and
vector potential in (\ref{eq5p10}) and (\ref{eq5p11}) are generated by
the polarizations terms associated with poloidally varying departures
from flux surfaces.
	
Next, we flux surface average using $\llangle \int
{d^3}v{f_0}\overline X Y\rrangle = \llangle \int {d^3}v{f_0}X\overline
Y \rrangle $, to obtain the flux surface averaged perpendicular \Amp{}
\begin{align}
\llangle {\tilde B_{||}}\rrangle - \tilde B_{||}^{(0)} \simeq
&\frac{{c{\beta _i}k_ \perp ^2{\rho _i}}}{{4{v_i}}}\left[(5\gamma
  {q^2}{\epsilon ^{ - 1/2}} + 3) \llangle \tilde \Phi \rrangle - 3
  {\tilde \Phi ^{(0)}}\right]\nonumber \\ + & \frac{{i q\beta {k_
  \perp }}}{2}\left\{ \frac{{5\gamma {\epsilon ^{1/2}}}}{2}\left[\llangle
  \Apat\rrangle - \tilde A_{||}^{(0)}\left(1 + \frac{{k_ \perp
      ^2{c^2}}}{{\omega _p^2}}\right)\right] + a \frac{3\eps }{2} 
\right\} \label{eq5p12},
\end{align}
where we used (\ref{A36}), (\ref{A49}), (\ref{A51}), (\ref{A52}), and
(\ref{A55}).  The $\Bz$ factors in (\ref{eq5p10}) must be treated with
care, even though $\ol{\vpa\cos\theta}=0$ for the trapped and
$\ol{\vpa\cos\theta}\sim \eps v$ for the passing, since they alter the
coefficient.  For now we retain the $\eps^{1/2}$ smaller $\llangle
\tilde \Phi \rrangle $ and ${\tilde \Phi ^{(0)}}$ terms compared to
the $\gamma {\epsilon ^{ - 1/2}}{q^2}\llangle \tilde \Phi \rrangle $
term.

In the parallel \Amp{} for now we keep $\ord(\eps^{3/2})$ corrections
to the typically large $\omega _p^2 / k_ \perp ^2{c^2}$ terms obtained
by using (\ref{A26}). By keeping the $\eps^{3/2}$ correction to
$\Apat$ terms we are making the easily satisfied assumption that
${\epsilon ^{3/2}}\omega _p^2 / k_ \perp ^2{c^2} \gg {q^2}\beta$, or
roughly $M/ m \gg \eps^{1/2} k_\perp ^2\rho _{pi}^2$, since we already
neglected the more complicated $Q^2$ corrections to the same
terms. The flux surface averaged parallel \Amp{} then becomes
\begin{align}
\llangle \Apat\rrangle & \left[1 + \frac{{\omega _p^2}}{{k_ \perp
      ^2{c^2}}}\left(1 - \gamma {\epsilon ^{3/2}}\right)\right] -
\tilde A_{||}^{(0)}\left[\left(1 + \frac{{\omega_p^2}}{{k_ \perp
      ^2{c^2}}}\right)(1 - \gamma {\epsilon ^{3/2}})\right] -a \left[
  \frac{\eps\op^2}{2\kpe^2c^2}(1-\gamma\eps^{3/2})+\Delta_S' \right]
\nonumber \\ \simeq &\frac{ iq\gamma {\epsilon ^{1/2}}\tilde
  B_{||}^{(0)}}{\kpe} + i\frac{{ c {\beta _i}q{k_ \perp }{\rho
      _i}}}{{4{v_i}}}\left[(\sigma {\epsilon ^{ - 1/2}}{q^2} + 5\gamma
  {\epsilon ^{1/2}} + 4\epsilon )\llangle \tilde \Phi \rrangle -
  4(\gamma {\epsilon ^{1/2}} + \epsilon ){\tilde \Phi
    ^{(0)}}\right],\label{eq5p13}
\end{align}
where the constant $\sigma\approx 5.3$ is evaluated in
Appendix~\ref{AppIntegrals}, and we make use of (\ref{A26}),
(\ref{A29}), (\ref{A31}), (\ref{A36}), (\ref{A56}) and (\ref{A63}).
We use (\ref{eq5p7}) and keep the Shafranov shift, $\Delta_S$, terms
from $\kpe\propto R\Bp \propto 1-\Delta_S'\cos\theta$ in the
$\langle\Apat\rangle$ and $\Aptz$ terms that arise from (\ref{eq5p2})
and the left hand side of (\ref{eq5p4}).  To simplify the flux surface
averaged parallel \Amp{} we ignored $\beta {\tilde B_{||}}$ and $\beta
\tilde B_{||}^{(0)}$ terms. Also, we have neglected all $b$ terms
because they are multiplied by $\beta$. Based on (\ref{rhresult}) we
must retain the $\sigma {q^2}{\epsilon ^{ - 1/2}}\llangle \tilde \Phi
\rrangle $ term in (\ref{eq5p13}), but will ignore the $\epsilon $ and
${\epsilon ^{3/2}}$ smaller term $(5\gamma {\epsilon ^{1/2}} +
4\epsilon )\llangle \tilde \Phi \rrangle $, as well as the $(\gamma
{\epsilon ^{1/2}} + \epsilon ){\tilde \Phi ^{(0)}}$ term, all
associated with the Bessel function corrections.
 
The terms $\gamma$ and $\sigma$ in (\ref{eq5p12}) and (\ref{eq5p13})
arise from the trapped particle responses to any initial perturbation,
while the $a \eps$ and $a\dsp$ terms are passing responses.  We
estimate the different behaviour of the passing and trapped (and barely
passing) by using $\overline {{v_{||}}} = 0$ and estimating
${v_{||}}\sim{\epsilon ^{1/2}}v$ and ${\int d ^3}v \propto {\epsilon
  ^{1/2}}$ for the trapped particles, while using $\overline
{{v_{||}}} \simeq {v_{||}}\left[1 + \ord(\epsilon )\right]$ for the
passing ones.

Next, we subtract the flux surface averaged equations from the full
equations. For the perpendicular \Amp{} we use (\ref{A49}), (\ref{A51}),
  (\ref{A52}), and (\ref{A55}) to find 
\begin{align}
b \simeq & \epsilon \tilde B_{||}^{(0)}
-\frac{iq\beta\kpe}{2}(3-\varsigma\eps^{1/2})\left[
  \langle\Apat\rangle-\Aptz\left(1+\frac{\kpe^2c^2}{\op^2}\right)
  \right] \nonumber \\ - & \frac{i\chi\eps^{3/2}q\beta\kpe
  a}{4}+\frac{c\varsigma\beta_i q^2\kpe^2\rhoi}{2\epsilon^{1/2}
  v_i}\langle\pht\rangle,
\label{eq5p14}
\end{align} 
where Bessel function terms are ignored since any $\cos\theta$ dependence is smaller by $\epsilon$, and the constants $\chi\approx 0.11$ and $\varsigma\approx 5.3$ are evaluated in Appendix~\ref{AppIntegrals}. The $\llangle
\Apat\rrangle $, $\tilde A_{||}^{(0)}$, and $\llangle \tilde \Phi
\rrangle$ terms are from passing contributions, while the ${\epsilon
  ^{3/2}}a$ term is a trapped contribution.  

To form the difference equation for the parallel \Amp{} we ignore
$\beta \llangle {\tilde B_{||}}\rrangle$, $\beta \tilde B_{||}^{(0)}$
and $\beta b$ terms. As a result, the poloidally varying parallel
\Amp{} reduces to
\begin{align} 
a\left(1 + \frac{{\epsilon^2 \omega _p^2}}{{2k_ \perp ^2{c^2}}}\right)
\simeq & \left(\frac{{\epsilon \omega _p^2}}{{k_ \perp ^2{c^2}}}+2\dsp
\right) \left(\llangle \Apat\rrangle - \tilde A_{||}^{(0)}\right)
-\eps \Aptz \label{eq5p15} \\+& \frac{iq\tilde B_{||}^{(0)}}{\kpe} +
i\frac{{ c {\beta _i}q{k_ \perp }{\rho _i}}}{{4{v_i}}}\left[(5\gamma
  {\epsilon ^{ - 1/2}}{q^2} + 4 )\llangle \tilde \Phi \rrangle -
  3{\tilde \Phi ^{(0)}}\right], \nonumber
\end{align}
where we make use of use (\ref{A26}), (\ref{A29}), (\ref{A31}),
(\ref{A36}), (\ref{A56}), and (\ref{A63}), and account for
$\kpe\propto 1-\dsp\cos\theta$. Only passing contributions enter for
$\llangle \Apat\rrangle$, $\tilde A_{||}^{(0)}$, $a$, and $\tilde
B_{||}^{(0)}$ terms, and only Bessel terms enter for ${\tilde \Phi
  ^{(0)}}$, while for $\llangle \tilde \Phi \rrangle$ both trapped and
Bessel contributions enter.

\subsection{ Parallel \Amp{} limit}
\label{sec:5p2}
Simulations are sometimes run ignoring the perpendicular \Amp. If we do the same by dropping all $b$, and $\tilde B_{||}^{(0)}$ terms in parallel \Amp, then (\ref{eq5p15}) becomes
\begin{equation} 
a\left(1 + \frac{{\epsilon^2 \omega _p^2}}{{2k_ \perp ^2{c^2}}}\right)
\simeq  \left(\frac{{\epsilon \omega _p^2}}{{k_ \perp ^2{c^2}}}+2\dsp
\right) \left(\llangle \Apat\rrangle - \tilde A_{||}^{(0)}\right)
-\eps \Aptz+ i\frac{{ c {\beta
      _i}q{k_ \perp }{\rho _i}}}{{2{v_i}}}\left[1 - ({\epsilon ^{1/2}}/
  2\gamma {q^2} )\right] {\tilde \Phi ^{(0)}},
\label{eq5p16}
\end{equation}
and (\ref{eq5p13}) reduces to 
\begin{align}
\llangle \Apat\rrangle \left[1 + \frac{{\omega _p^2}}{{k_ \perp ^2{c^2}}}\left(1
  - \gamma {\epsilon ^{3/2}}\right)\right] - & \tilde A_{||}^{(0)}\left[\left(1 +
  \frac{{\omega _p^2}}{{k_ \perp ^2{c^2}}}\right)(1 - \gamma {\epsilon
    ^{3/2}})\right] -a \left[
  \frac{\eps\op^2}{2\kpe^2c^2}(1-\gamma\eps^{3/2})+\Delta_S' \right]\nonumber \\ 
\simeq &
i\frac{{ c\sigma  {\beta _i}q{k_ \perp }{\rho _i}}}{{4\gamma {v_i}}}\left[1
  - ( {\epsilon ^{1/2}}/ \gamma {q^2})  - (4{\gamma ^2}{\epsilon
    ^{1/2}}/ \sigma )\right] {\tilde \Phi ^{(0)}},
\label{eq5p17}
\end{align}
where we have used $\beta \ll 1$ so we can use (\ref{rhresult}) to
eliminate $\llangle \tilde \Phi \rrangle$.

If $\beta\rightarrow 0$ at finite $\kpe$ so that $\omega _p^2 / k_
\perp ^2{c^2} \to   0$ then (\ref{eq5p16}) and (\ref{eq5p17}) reduce
to the large skin depth results
\begin{equation}
\frac{{{v_i}}}{c}\llangle \Apat\rrangle \to \frac{{{v_i}}}{c}\tilde
A_{||}^{(0)}(1 - \gamma {\epsilon ^{3/2}}) + i\frac{{\sigma  {\beta
      _i}q{k_ \perp }{\rho _i}}}{{4\gamma }}{\tilde \Phi ^{(0)}}
\label{eq5p18} 
\end{equation}
and
\begin{equation}
\frac{{{v_i}}}{c}a \to -
              \frac{{{v_i}}}{c}\epsilon \tilde A_{||}^{(0)} +
              i\frac{{{\beta _i}q{k_ \perp }{\rho _i}}}{2}{\tilde \Phi
                ^{(0)}},
\label{eq5p19} 
\end{equation}
 where we neglect ${\epsilon ^{1/2}}$ corrections to ${\tilde \Phi
   ^{(0)}}$ terms. Interestingly, for $\tilde A_{||}^{(0)} = 0$ we see
 that $\llangle \Apat\rrangle /a \simeq \sigma /2\gamma \simeq \ord(1)$,
 meaning that substantial poloidal variation occurs when ${\tilde \Phi
   ^{(0)}} \ne 0$. When ${\tilde \Phi ^{(0)}} = 0$, $\llangle
 \Apat\rrangle \simeq \tilde A_{||}^{(0)}(1 - \gamma {\epsilon
   ^{3/2}})$ and $a \simeq - \epsilon \llangle \Apat\rrangle$, and
 poloidal variation is somewhat weak.

More interestingly, we consider finite $\beta\ll 1$ by allowing
$\epsilon ^{3/2} \gg k_ \perp ^2{c^2}/ \omega
_p^2 $ or ${q^2}{\beta _i}\eps^{1/2}\gg Zm k_ \perp ^2\rho
_{pi}^2 / M$.  Continuing to neglect ${\epsilon ^{1/2}}$ corrections
in this small skin depth limit we find
\begin{equation}
\frac{{{v_i}}}{c}\llangle
 \Apat\rrangle \simeq \frac{\vi}{c}\Aptz\left(1-\frac{\eps^2}{2}\right)
+\frac{i\eps \beta_i \kpe \rho_{pi}}{4}{\tilde \Phi }^{(0)},
\label{eq5p20} 
\end{equation}
and
\begin{equation}
\frac{{{v_i}}}{c}a \simeq -\frac{\vi}{c}\eps\Aptz
+\frac{i\eps \beta_i \kpe \rho_{pi}}{2}{\tilde \Phi }^{(0)}.
\label{eq5p21} 
\end{equation}
 When ${\tilde \Phi ^{(0)}} = 0$, $\llangle \Apat\rrangle \simeq
 \tilde A_{||}^{(0)} \simeq  -  a/\epsilon$ so only weak poloidal variation
 occurs. However, for $\tilde A_{||}^{(0)} \simeq 0$, we find
 that $a/\llangle \Apat\rrangle \simeq 2$,
 which results in strong poloidal variation.  

In both the small and large skin depth limits, when ${\tilde \Phi
  ^{(0)}} = 0$, only small changes in $\Apat$ occur since $\Apat \simeq
\llangle \Apat\rrangle \simeq \tilde A_{||}^{(0)} \gg a$. However, for
$\tilde A_{||}^{(0)} \simeq 0$, we see that poloidal variation in
$\Apat$ arises due to ${\tilde \Phi ^{(0)}}$ with $a/\llangle
\Apat\rrangle \sim 1$ in both limits.

\subsection{Full \Amp}
\label{sec5p3}
The general case retains the perpendicular \Amp. Using (\ref{eq5p13}) and (\ref{eq5p15}) with (\ref{rhresult}) inserted and ignoring ${\epsilon ^{1/2}}$ corrections we find
\begin{equation}
a\left(1 + \frac{{\epsilon^2 \omega _p^2}}{{2k_ \perp ^2{c^2}}}\right)
\simeq  \left(\frac{{\epsilon \omega _p^2}}{{k_ \perp ^2{c^2}}}+2\dsp
\right) \left(\llangle \Apat\rrangle - \tilde A_{||}^{(0)}\right)
-\eps \Aptz + i\frac{{ c {\beta
      _i}q{k_ \perp }{\rho _i}}}{{2{v_i}}} {\tilde \Phi ^{(0)}} +
iqk_ \perp ^{ - 1}\tilde B_{||}^{(0)}
\label{eq5p22} 
\end{equation}
and 
\begin{equation}
\left(\llangle \Apat\rrangle  - \tilde A_{||}^{(0)}\right)\left(1 + \frac{{\omega _p^2}}{{k_ \perp ^2{c^2}}}\right) -a \left[
  \frac{\eps\op^2}{2\kpe^2c^2}(1-\gamma\eps^{3/2})+\Delta_S' \right] \simeq i\frac{{  c\sigma  {\beta _i}q{k_ \perp }{\rho _i}}}{{4\gamma {v_i}}} {\tilde \Phi ^{(0)}} + iq\gamma {\epsilon ^{1/2}}k_ \perp ^{ - 1}\tilde B_{||}^{(0)}.
\label{eq5p23} 
\end{equation}
In addition, (\ref{eq5p12}) and (\ref{eq5p14}) reduce to 
\begin{align}
\llangle {\tilde B_{||}}\rrangle   - & \tilde B_{||}^{(0)} \simeq
\frac{{c{\beta _i}k_ \perp ^2{\rho _i}}}{{2{v_i}}} {\tilde \Phi
  ^{(0)}} 
\label{eq5p24}  \\
+ &\frac{{iq\beta }k_ \perp}{2}\left\{ \frac{{5\gamma {\epsilon
      ^{1/2}}}}{2}\left[\llangle \Apat\rrangle - \tilde
  A_{||}^{(0)}\left(1 + \frac{{k_ \perp ^2{c^2}}}{{\omega
      _p^2}}\right)\right] + a \frac{3\eps}{2} \right\} \nonumber
\end{align}
 and
\begin{align}
b \simeq & \epsilon \tilde B_{||}^{(0)}
-\frac{iq\beta\kpe}{2}(3-\varsigma\eps^{1/2})\left[
  \langle\Apat\rangle-\Aptz\left(1+\frac{\kpe^2c^2}{\op^2}\right)
  \right] \nonumber \\ - & \frac{i\chi\eps^{3/2}q\beta\kpe
  a}{4}+\frac{c\varsigma\beta_i \kpe^2\rhoi}{2\gamma v_i}\phtz,
\label{eq5p25}
\end{align} 
 where we used the lowest order version of (\ref{rhresult}),
\begin{equation}
\llangle \tilde \Phi \rrangle   \simeq \frac{{\epsilon
  ^{1/2}}{\tilde \Phi ^{(0)}} }{ \gamma {q^2}}.
\label{eq5p26} 
\end{equation}
 
 When $\tilde B_{||}^{(0)} = 0$ we may employ
 (\ref{eq5p18})-(\ref{eq5p21}) to find $\llangle {\tilde
   B_{||}}\rrangle$ and $b$ from (\ref{eq5p24}) and (\ref{eq5p25}). In
 the $\omega _p^2 / k_ \perp ^2{c^2} \to 0$ limit (\ref{eq5p18}) and
 (\ref{eq5p19}) are used along with (\ref{eq5p24}) and (\ref{eq5p25})
 to find
\begin{equation}
\frac{{{v_i}\llangle
     {{\tilde B}_{||}}\rrangle }}{{{k_ \perp }c}}  \to \frac{{{\beta
       _i}{k_ \perp }{\rho _i}}}{2} {\tilde \Phi ^{(0)}} -
 i\frac{{5\gamma {\epsilon ^{1/2}}q\beta k_ \perp ^2{c^2}}}{{4\omega
     _p^2}}\frac{{{v_i}}}{c}\tilde A_{||}^{(0)}
\label{eq5p27} 
\end{equation}
 and
\begin{equation} 
\frac{{{v_i}b}}{{{k_ \perp }c}}   \to    iq\beta \frac{{3k_ \perp ^2{c^2}}}{{2\omega _p^2}}\frac{{{v_i}}}{c}\tilde A_{||}^{(0)} + \frac{\varsigma{{\beta _i}{k_ \perp }{\rho _i}}}{{2\gamma }}{\tilde \Phi ^{(0)}}.
\label{eq5p28} 
\end{equation}
Consequently, when ${\tilde \Phi ^{(0)}}  = 0$ we see that $\llangle
{\tilde B_{||}}\rrangle  / b \simeq - 5\gamma {\epsilon ^{1/2}}/ 6$
so strong poloidal variation occurs in ${\tilde B_{||}}$. For $\tilde
A_{||}^{(0)}  = 0$ the poloidal variation of ${\tilde B_{||}}$ is
strong since $b/\llangle {\tilde B_{||}}\rrangle   \simeq  {\varsigma}/ \gamma \approx 3.2$.
	
For $\Bpatz=0$, $\beta \ll 1$, and $\epsilon ^{3/2} \gg k_ \perp
^2{c^2}/ \omega _p^2$ we use (\ref{eq5p20}) and (\ref{eq5p21}) to find
the small skin depth forms
\begin{equation}
\frac{{{v_i}\llangle {{\tilde B}_{||}}\rrangle }}{{{k_ \perp }c}} 
\simeq \frac{{\epsilon {\beta _i}{k_ \perp }{\rho _{pi}}}}{{2q}}
       {\tilde \Phi ^{(0)}} - \frac{{i3\epsilon^2 \beta q {v_i}}}{4c}\tilde
           A_{||}^{(0)}
\label{eq5p29} 
\end{equation}
and
\begin{equation}
\frac{{{v_i}b}}{{{k_ \perp }c}} \simeq\frac{{i3 \beta q
    {v_i}}}{4c}\left(\eps^2+\frac{2\kpe^2c^2}{\op^2}\right)\tilde A_{||}^{(0)} + \frac{{{\varsigma\epsilon}{\beta
      _i}{k_ \perp }{\rho _{pi}}}}{{2\gamma q}}{\tilde \Phi ^{(0)}}.
\label{eq5p30} 
\end{equation}
When $\tilde A_{||}^{(0)} = 0$ the poloidal variation of ${\tilde
  B_{||}}$ is strong since $b/\llangle {\tilde B_{||}}\rrangle
\simeq{\varsigma}/\gamma\approx 3.2$. For ${\tilde \Phi ^{(0)}} = 0$
it is also strong with $b/\llangle {\tilde B_{||}}\rrangle \simeq -
[1+(2\kpe^2c^2/\eps^2\op^2)]$. Consequently, in the small skin depth
limit, for $\tilde B_{||}^{(0)} = 0$ the poloidal variation of
${\tilde B_{||}}$ is strong when $\tilde A_{||}^{(0)} = 0$ and weak
when ${\tilde \Phi ^{(0)}} = 0$.

Next we consider the case ${\tilde \Phi ^{(0)}}  = 0$ and $\tilde A_{||}^{(0)}  = 0$ when $\tilde B_{||}^{(0)} \ne  0$ by using (\ref{eq5p22})-(\ref{eq5p25}). The $\omega _p^2  /  k_ \perp ^2{c^2}   \to   0$ limit gives
\begin{equation} 
{k_ \perp }a \to   iq\tilde B_{||}^{(0)},
\label{eq5p31} 
\end{equation}
\begin{equation} 
{k_ \perp }\llangle \Apat\rrangle \to   iq\gamma {\epsilon
  ^{1/2}}\tilde B_{||}^{(0)}.
\label{eq5p32} 
\end{equation}
\begin{equation}
\llangle {\tilde B_{||}}\rrangle   \to   \left[1 - (3 + 5{\gamma
    ^2})(\epsilon {q^2}\beta / 4)\right] \tilde B_{||}^{(0)},
\label{eq5p33} 
\end{equation}
and
\begin{equation}
b   \to  (\epsilon + 3\gamma {\epsilon ^{1/2}}{q^2}\beta/2 ) \tilde
B_{||}^{(0)} .
\label{eq5p34} 
\end{equation}
Consequently, the poloidal variation of ${\tilde B_{||}}$ is weak
(with $b /\llangle {\tilde B_{||}}\rrangle   \simeq  \epsilon$),
while the poloidal variation of $\Apat$ is strong (with $\llangle
\Apat\rrangle /a \simeq  \gamma {\epsilon ^{1/2}}$) in this large
skin depth limit.

In the small skin depth limit $\epsilon^{3/2}\gg k_ \perp ^2{c^2}/
\omega _p^2$ for finite $\beta\ll 1$ and $\phtz=0=\Aptz$ we use
(\ref{eq5p22}) and (\ref{eq5p23}) to find $\eps a \simeq 2(1+\gamma
\eps^{3/2})\langle\Apat\rangle$. Then (\ref{eq5p22}) to (\ref{eq5p25})
give
\begin{equation}
{k_ \perp }a \simeq \frac{iq\tilde
  B_{||}^{(0)}}{1+(\gamma\eps^{7/2}\op^2/2\kpe^2c^2)},
\label{eq5p35} 
\end{equation}
\begin{equation}
{k_ \perp }\llangle \Apat\rrangle  \simeq   \frac{i\eps q \tilde
  B_{||}^{(0)}}{2+(\gamma\eps^{7/2}\op^2/\kpe^2c^2)}.
\label{eq5p36} 
\end{equation}
\begin{equation}
\llangle {\tilde B_{||}}\rrangle \simeq \tilde B_{||}^{(0)} \left\{1-
\frac{3 \eps \beta q^2
}{2[2+(\gamma\eps^{7/2}\op^2/\kpe^2c^2)]}\right\},
\label{eq5p37} 
\end{equation}
and
\begin{equation}
b   \simeq  \tilde B_{||}^{(0)} \left\{\eps+
\frac{3 \eps \beta q^2
}{2[2+(\gamma\eps^{7/2}\op^2/\kpe^2c^2)]}\right\}.
\label{eq5p38} 
\end{equation}
As a result, for this $\tilde B_{||}^{(0)} \ne 0$ case with ${\tilde
  \Phi ^{(0)}} = 0$ and $\tilde A_{||}^{(0)} = 0$, the poloidal
variation of $\Apat$ is strong with $\llangle \Apat\rrangle / a \simeq
\eps / 2$, while the poloidal variation of ${\tilde B_{||}}$ varies
from weak to strong with $b/\llangle {\tilde B_{||}}\rrangle \simeq
\epsilon$ to $1$.

\subsection{Quasineutrality}
\label{sec:qn}
To complete our treatment of the Maxwell equations we need to form
quasineutrality with its finite $\beta$ effects retained using
\begin{equation}
\tilde \Phi \sum \frac{Z^2 e^2 n}{T} = \sum Ze\int {{d^3}} v\overline
       {{{\tilde h}_ \ast }(t)}  {e^{i(L - Q)}} = \sum Ze\int {{d^3}}
       v\overline {{{\tilde h}_ \ast }(t)}  {J_0}{e^{ - iQ}}.
\label{eq5p39} 
\end{equation}

After inserting (\ref{intgk}) and (\ref{initialaverage}), we only
retain Bessel function modifications to $\tilde \Phi \simeq \llangle
\tilde \Phi \rrangle$ and ${\tilde \Phi ^{(0)}}$ terms. We then find
\begin{align}
\tilde \Phi  \sum  \frac{Z^2 e^2 n}{T} = &\left(\llangle \tilde \Phi \rrangle -
       {\tilde \Phi ^{(0)}}\right)\sum \frac{Z^2 e^2}{T}\int {{d^3}}
       v{f_0}{J_0}{e^{ - iQ}}\overline {{J_0}{e^{iQ}}} \nonumber 
\\ + & {\tilde \Phi ^{(0)}}\sum \frac{Z^2 e^2}{T}\int {{d^3}}
       v{f_0}H{J_0}{e^{ - iQ}}\overline {({e^{iQ}}/
         {J_0})} \label{eq5p40} 
\\ - & \sum \frac{Z^2 e^2}{cT}\int
       {{d^3}} v{f_0}{e^{ - iQ}}\left[\overline {{{\tilde
               A}_{||}}{v_{||}}{e^{iQ}}} - \tilde A_{||}^{(0)}(1 + k_
         \perp ^2{c^2}/ \omega _p^2)\overline {{v_{||}}{e^{iQ}}} \right]\nonumber
\\- &\tilde B_{||}^{(0)}
\frac{2{\Bzl}}{\beta B_0^2}\sum Ze\int {{d^3}} v{f_0}{e^{ -
     iQ}}\overline {{e^{iQ}}} + \sum \frac{ZeM}{  2T{B_0}}\int
 {{d^3}} v{f_0}v_ \bot ^2{e^{ - iQ}}(\overline {{{\tilde
       B}_{||}}{e^{iQ}}} - \tilde B_{||}^{(0)}\overline {{e^{iQ}}} ).
 \nonumber
\end{align}

Expanding the Bessel functions and the exponentials in $Q$, inserting $H$
with $\alpha = k_ \perp ^2T / 2M{\Omega ^2}$, and using unperturbed
quasineutrality leaves 
\begin{align} 
0 = & \llangle \tilde \Phi \rrangle \sum \frac{Z^2e^2}{ T}\int d^3
vf_0\left[Q\bar Q - \frac{Q^2 + \overline {Q^2}}{ 2}\right] -
\left(\llangle \tilde \Phi \rrangle - {\tilde \Phi ^{(0)}}\right) \sum
\frac{{Z^2}{e^2}nk_\perp^2 }{ M{\Omega ^2}}\nonumber \\ - & i\sum
\frac{Z^2e^2}{2cT}\int d^3 vf_0\left[\overline {{{\tilde
        A}_{||}}{v_{||}}Q } {Q^2} - \overline {{{\tilde
        A}_{||}}{v_{||}}{Q^2} } Q - \frac{\overline {{{\tilde
          A}_{||}}{v_{||}}} {Q^3} - \overline {{{\tilde
          A}_{||}}{v_{||}}{Q^3}} }{3} \right]\nonumber \\ + & \tilde
A_{||}^{(0)} \left(1 + \frac{k_\perp^2c^2}{\omega_p^2}\right) i\sum
\frac{Z^2e^2} { 2cT}\int {{d^3}} v{f_0}\left[\overline {{v_{||}}Q }
  {Q^2} - \overline {{v_{||}}{Q^2} } Q - \frac{\overline {{v_{||}}}
    {Q^3} - \overline {{v_{||}}{Q^3}} }{3}\right]\nonumber
\\ - & \tilde
B_{||}^{(0)}\frac{2{\Bzl}}{\beta B_0^2}\sum Ze\int {{d^3}} v{f_0}\left[Q\bar Q -
  ({Q^2} + \overline {{Q^2}} )  / 
  2\right] \label{eq5p41} 
\\ + & \sum \frac{ZeM}{ 2T{B_0}} \int {{d^3}} v{f_0}v_
\perp ^2\left\{ \left[Q\overline {{{\tilde B}_{||}}Q} - \frac{{Q^2}\overline
  {{{\tilde B}_{||}}} + \overline {{Q^2}{{\tilde B}_{||}}} }{2}\right] -
  \tilde B_{||}^{(0)}\left[Q\bar Q - \frac{{Q^2} + \overline {{Q^2}}}{ 2}\right]\right\}
  .\nonumber
\end{align}

We have already shown that the poloidal variation of $\tilde \Phi$ is
negligible so we only require the flux surface average of
  (\ref{eq5p41}), which is
\begin{align}
\llangle \tilde \Phi \rrangle & \left[\sum \frac{{Z^2}{e^2} }{ T}\llangle \int
    {{d^3}} v{f_0}Q(Q - \bar Q)\rrangle + \sum \frac{{Z^2}{e^2}nk_ \perp ^2
    }{ M{\Omega ^2}}\right] =   {\tilde \Phi ^{(0)}}\sum \frac{{Z^2}{e^2}nk_
  \perp ^2 }{ M{\Omega ^2}}\nonumber 
\\ -& i\sum \frac{{Z^2}{e^2} }{
  2cT}\llangle \left[\Apat - \tilde A_{||}^{(0)}\left(1 + \frac{k_ \perp ^2{c^2}}{
    \omega _p^2}\right)\right]\int {{d^3}} v{f_0}{v_{||}}\left[\overline {{Q^2} } Q -
    \overline {Q } {Q^2} + \frac{{Q^3} - \overline {{Q^3}}}{3}\right]\rrangle
  \nonumber 
\\ +& \tilde
  B_{||}^{(0)}\frac{2{\Bzl}}{ \beta B_0^2}\Sigma Ze\langle \int {{d^3}}
  v{f_0}Q(Q - \bar Q)  \rangle,
\label{eq5p42}
\end{align}	
where we neglect $\beta \llangle {\tilde B_{||}}\rrangle$, $\beta
\tilde B_{||}^{(0)}$ and $\beta b$ corrections.

Performing the integrals using (\ref{A26}), (\ref{A59}),
(\ref{A62}), (\ref{A65}), and (\ref{A66}), and noting that only the
trapped ion and Bessel contributions matter we obtain
\begin{align}
(\gamma & {q^2}{\epsilon ^{ - 1/2}} +  1)\llangle \tilde \Phi \rrangle =
    {\tilde \Phi ^{(0)}} + \gamma {\epsilon ^{1/2}}q{\rho
    _{pi}}\frac{{v_i}}{ c\beta }\tilde B_{||}^{(0)}\label{eq5p43} 
\\ - & i\frac{{v_i}}{4c}{\epsilon ^{1/2}}{q^2}{k_ \perp }{\rho _{pi}}\left\{ \sigma
   \left[\llangle \Apat\rrangle - \tilde A_{||}^{(0)}\left(1 + \frac{k_ \perp
    ^2{c^2}}{ \omega _p^2}\right)\right] + \frac{5\gamma a }{ 2}\right\} 
\nonumber.
\end{align}

 	In the limit in which we take $\beta\rightarrow 0$ at finite $\kpe$ (such that $\omega _p^2  /  k_ \perp ^2{c^2}   \to   0$), (\ref{eq5p43}) with (\ref{eq5p19}) and (\ref{eq5p20}) inserted gives the large skin depth expression
\begin{equation} 
\llangle \tilde \Phi \rrangle \to   \left(\frac{\gamma {q^2}}{\epsilon ^{1/2}} +
    1\right)^{-1}  \left[{{\tilde \Phi }^{(0)}} +
      i\frac{{v_i}\tilde A_{||}^{(0)}}{ 4c}\frac{k_ \perp ^2{c^2}}{ \omega
      _p^2}(\sigma  {\epsilon ^{1/2}}{q^2}{k_ \perp }{\rho _{pi}}) +
      \frac{\gamma {\epsilon ^{1/2}}}{ \beta }\frac{q {\rho _{pi}}{v_i}\tilde
      B_{||}^{(0)}}{ c }\right]\label{eq5p44}.
\end{equation}
	
In the more interesting finite limit, for which $\epsilon ^{3/2} \gg
k_ \perp ^2{c^2}/ \omega _p^2$, using (\ref{eq5p20}), (\ref{eq5p21}),
(\ref{eq5p35}), and (\ref{eq5p36}) we find the small skin depth result
\begin{align} 
\llangle \tilde \Phi \rrangle \simeq & \frac{{{{\tilde \Phi
      }^{(0)}}}}{{(\gamma {q^2}{\epsilon ^{ - 1/2}} + 1)}} \left[1 +
  \frac{\sigma+5\gamma}{16}\beta_i\eps^{3/2}q^2\kpe^2\rho_{pi}^2\right] \label{eq5p45}
\\ +& \frac{i5 \gamma \eps^{3/2}q^2\kpe\rho_{pi}\vi \Aptz}{8(\gamma
  q^2\eps^{-1/2}+1)c}+ \frac{\gamma
  \eps^{1/2}q\kpe\rho_{pi}\vi\Bpatz}{(\gamma
  q^2\eps^{-1/2}+1)\beta\kpe c}\nonumber.
\end{align}
Clearly, the modification of the electrostatic limit is small when
$\tilde B_{||}^{(0)} = 0 = \tilde A_{||}^{(0)}$. From (\ref{eq5p45}),
we see that initial magnetic perturbations as large as ${\epsilon
  ^{1/2}}q{\rho _{pi}}{v_i}\tilde B_{||}^{(0)}/ c\sim\beta {\tilde
  \Phi ^{(0)}}$ and/or ${\epsilon ^{3/2}} {q^2} {k_ \perp }{\rho
  _{pi}}{v_i}\tilde A_{||}^{(0)}/ c\sim{\tilde \Phi ^{(0)}}$ are
required to obtain order unity corrections to the electrostatic
response. For the same size initial perturbations in the expression
(\ref{eq5p44}) the response to $\tilde B_{||}^{(0)}$ remains the same,
but that due to $\tilde A_{||}^{(0)}$ is very small.

%_________________________________________________
%-------------------------------------------------
%_________________________________________________

\section{Zonal flow responses in terms of initial field values: 15 test cases}
\label{sec:plots}  
In this section we present a summary of our results in the small skin
depth limit. Before doing so we note that
\begin{equation}
\frac{{k_ \perp ^2{c^2}}}{{\omega _p^2}} = \frac{{Z(1 + \tau
    ){\epsilon ^2}m}}{{{q^2}\beta M}}k_ \perp ^2\rho _{pi}^2 =
\frac{{Z(1 + \tau )m}}{{\beta M}}k_ \perp ^2\rho _i^2 = \frac{{(1 +
    \tau )}}{{\tau  \beta }}k_ \perp ^2\rho _e^2,
\label{eq6p1}
\end{equation}
 with $\beta = {\beta _i}(1 + \tau )$ and $\tau = Z 
 {T_e}  /  {T_i}$. We, of course, neglect mass
 ratio corrections in the plasma frequency.

The large skin depth limit of $\beta\rightarrow 0$ requires
${q^2}\beta \ll Zmk_ \perp ^2\rho _{pi}^2 /M \ll Zm/M$ since we must
keep $k_ \perp ^2\rho _{pi}^2 \ll 1$ for our analysis to
hold. Consequently, the more interesting limit is that of small skin
depth ${\epsilon ^2}\sim\epsilon {\Delta '_S}\gg k_ \perp ^2{c^2}/
\omega _p^2 $ and small ${\beta _i}$ ($1\gg \beta_i\gtrsim
Zm\kpe^2\rho_{pi}^2/q^2M$), which will be our focus. In the following
we summarize our results for this limit.  We will find that
electromagnetic perturbations give an electrostatic potential response
comparable to the electrostatic limit of $\llangle \tilde \Phi
\rrangle \simeq {\epsilon ^{1/2}}{\tilde \Phi ^{(0)}} / \gamma {q^2}$
when ${v_i}\Aptz{\epsilon ^{3/2}}{q^2}{k_ \perp }{\rho _{pi}}/
c\sim{\tilde \Phi ^{(0)}}$ or ${\epsilon ^{1/2}}q {\rho
  _{pi}}{v_i}\tilde B_{||}^{(0)}/ c\sim\beta {\tilde \Phi ^{(0)}}$.

\subsection{Electromagnetic response to an electrostatic initial perturbation ($\tilde B_{||}^{(0)}   = 0 = \Aptz$)}
\label{sec:elstat}
 In this case $\llangle \tilde \Phi \rrangle / {\tilde \Phi ^{(0)}}$
 has a very small linear in $\beta$ correction to the electrostatic
 Rosenbluth and Hinton form as seen from (\ref{eq5p45}),
 \begin{equation}
\left(\frac{\gamma {q^2}}{\epsilon ^{1/2}} + 1\right)\frac{{\llangle \tilde
     \Phi \rrangle }}{{{{\tilde \Phi }^{(0)}}}} \simeq 1 +
  \frac{\sigma+5\gamma}{16}\beta_i\eps^{3/2}q^2\kpe^2\rho_{pi}^2,
\label{eq6p2}
\end{equation}
 where $\sigma\approx 5.3$. Moreover, $\llangle \Bpat\rrangle {\rho
   _i}{v_i}  /  c{\tilde \Phi ^{(0)}}$ and $b{\rho
   _i}{v_i}  /  c{\tilde \Phi ^{(0)}}$ are linear
 in $\beta$ and are seen from (\ref{eq5p29}) and (\ref{eq5p30}) to be
 given by
\begin{equation}
\frac{{{v_i}\llangle {{\tilde
         B}_{||}}\rrangle }}{{{k_ \perp }c{{\tilde \Phi
       }^{(0)}}}}   \simeq \frac{{\epsilon {\beta _i}{k_
       \perp }{\rho _{pi}}}}{{2q}}    
\label{eq6p3}
\end{equation}
 and
 \begin{equation}
\frac{{{v_i}b}}{{{k_ \perp }c{{\tilde \Phi }^{(0)}}}}  
 \simeq  \frac{{{\varsigma \epsilon}{\beta _i}{k_ \perp }{\rho
       _{pi}}}}{{2\gamma q}}.
\label{eq6p4}
\end{equation} 
 From (\ref{eq5p20}) and (\ref{eq5p21}) we find
 \begin{equation}
\frac{{{v_i}\llangle {{\tilde A}_{||}}\rrangle }}{{ic{{\tilde
         \Phi }^{(0)}}}} \simeq \frac{\eps\beta_i\kpe\rho_{pi}}{4},
\label{eq6p5}
\end{equation}
 and
\begin{equation}
\frac{{{v_i}a}}{{ic{{\tilde \Phi }^{(0)}}}} \simeq \frac{{\epsilon
    {\beta _i}{k_ \perp }{\rho _{pi}}}}{2},
\label{eq6p6}
\end{equation}
where all electromagnetic responses are proportional to $\beta$. The
correction to the electrostatic result is very small, but all
electromagnetic responses are larger and of the same order so strong
poloidal variation occurs.

\subsection{Electromagnetic response to a plucked field line initial condition ($\tilde B_{||}^{(0)}   = 0 = {\tilde \Phi ^{(0)}}$)}
\label{sec:plucked}

Using (\ref{eq5p45}) gives the electrostatic potential response for
this $\Aptz  \ne 0$ case to be
\begin{equation}
\frac{{c\llangle \tilde \Phi \rrangle }}{{i{v_i}\Aptz}} \simeq
\frac{{5\gamma \epsilon ^{3/2} q^2 {k_ \perp }{\rho _{pi}}}}{8(\gamma
  q^2\eps^{-1/2}+1)} \simeq \frac{{5{\epsilon ^2}{k_ \perp }{\rho
      _{pi}}}}{8}.
\label{eq6p7}
\end{equation}  
For this case, the parallel vector potential responses from
(\ref{eq5p20}) and (\ref{eq5p21}) are given by
\begin{equation}
\frac{{\llangle {{\tilde A}_{||}}\rrangle }}{{\Aptz}} \simeq 1 -
\frac{{\epsilon^2}}{2} 
\label{eq6p8}
\end{equation} 
and
\begin{equation}
\frac{a}{{\Aptz}} \simeq - \epsilon,
\label{eq6p9}
\end{equation} 
showing that $\Apat$ will only be very slightly perturbed from $\Aptz$
with weak poloidal variation due to field line plucking satisfying $R
\Apat\simeq R_0(\psi) \Aptz$. The $\beta$ proportional $\Bpat$
responses from (\ref{eq5p29}) and (\ref{eq5p30}) give $b$ and
$\llangle {{\tilde B}_{||}}\rrangle$ as comparable as seen from
\begin{equation}
\frac{{i \llangle {{\tilde B}_{||}}\rrangle }}{{{k_ \perp
    }\Aptz}}   \simeq 
\frac{{3\epsilon^2 \beta q}}{4}
\label{eq6p10}
\end{equation} 
and
\begin{equation}
\frac{b}{{i{k_ \perp }\Aptz}} \simeq \frac{{3  \beta q}}{4}\left(\eps^2+\frac{2\kpe^2 c^2}{\op^2} \right).
\label{eq6p11}
\end{equation}

\subsection{Electromagnetic response to a compressed field line 
initial condition ($\Aptz = 0  =  {\tilde \Phi ^{(0)}}$)}
\label{sec:compressed}
The response $c{k_ \perp }\llangle \tilde \Phi \rrangle / 
{v_i}\tilde B_{||}^{(0)}$ is proportional to $1/\beta$ as seen from
(\ref{eq5p45}) so in this $\tilde B_{||}^{(0)} \ne 0$ case we multiply
through by $\beta$ to form
\begin{equation}
\frac{{c\beta {k_ \perp }\llangle \tilde \Phi \rrangle }}{{{v_i}\tilde
    B_{||}^{(0)}}} \simeq \frac{\gamma
  \eps^{1/2}q\kpe\rho_{pi}}{\gamma q^2\eps^{-1/2}+1} \simeq
\frac{\eps\kpe\rho_{pi}}{q}.
\label{eq6p12}
\end{equation}
The responses $\llangle \Bpat\rrangle /\tilde B_{||}^{(0)}$ and
$b /\tilde B_{||}^{(0)}$ are given by (\ref{eq5p37}) and (\ref{eq5p38}) for this
field line stretching case:
\begin{equation}
\frac{{\llangle {{\tilde B}_{||}}\rrangle }}{{\tilde B_{||}^{(0)}}} 
\simeq  1 - \frac{3\eps q^2\beta}{2[2+(\gamma\eps^{7/2}\op^2/\kpe^2c^2)]} \simeq 1
\label{eq6p13}
\end{equation}
and 
\begin{equation}
\frac{b}{{\tilde B_{||}^{(0)}}}   \simeq  \eps + \frac{3 \eps  q^2\beta}{2[2+(\gamma\eps^{7/2}\op^2/\kpe^2c^2)]} \simeq \eps.
\label{eq6p14}
\end{equation}
The responses ${k_ \perp }\llangle \Apat\rrangle /  i\tilde
B_{||}^{(0)}$ and ${k_ \perp }a    / i 
  \tilde B_{||}^{(0)}$ are obtained from (\ref{eq5p35}) and
(\ref{eq5p36})
\begin{equation}
\frac{{{k_ \perp }\llangle {{\tilde A}_{||}}\rrangle }}{{i\tilde
    B_{||}^{(0)}}} \simeq \frac{\eps
  q}{2+(\gamma\eps^{7/2}\op^2/\kpe^2c^2)}
\label{eq6p15}
\end{equation}
and
\begin{equation}
\frac{{{k_ \perp }a}}{{i\tilde B_{||}^{(0)}}} \simeq
\frac{q}{1+(\gamma\eps^{7/2}\op^2/2\kpe^2c^2)}.
\label{eq6p16}
\end{equation}
Stretching or compressing field lines causes very strong poloidal
variation in $\Apat$, but weak poloidal variation in $\Bpat$
satisfying $R_0(\psi)\Bpat\simeq R \Bpatz $.

\section{Conclusions}
\label{conclusions}

We have derived approximate analytical expressions for long
wavelength, collisionless zonal flow residual responses at low $\beta$
in the Shafranov shifted, circular cross section, large aspect ratio
limit of a tokamak. To do so, we formulate and solve a Maxwell-Vlasov
description in the form of an initial value problem, retaining the
fully self-consistent spatial perturbations in the electric and
magnetic fields. As zonal flow perturbations are axisymmetric, the
only magnetic field topology change allowed is the switch between
rational and irrational field lines within a flux surface - no
magnetic islands can be formed. The choice of the initial condition in
the non-adiabatic part of the distribution function must be consistent
with Maxwell's equations, but it is otherwise arbitrary. The specific
choice we make is motivated by the desire to recover the usual long
wavelength result in the electrostatic limit that has the residual
proportional to the ratio of the classical polarization over the
classical plus neoclassical polarization. Also, our choice of initial
conditions is such that at t = 0 the initial electrostatic, and shear-
and compressional magnetic perturbations contribute only to
quasineutrality, the parallel and the perpendicular components of
\Amp, respectively. This form is convenient since then the initial
conditions in the amplitude of the fields can be chosen independently.
	
The results we obtain are expected to prove useful for testing
turbulent electromagnetic gyrokinetic codes just as Rosenbluth and
Hinton has proven useful in the electrostatic limit. The
electromagnetic case is of course far more complex, and further
complicated by the fact that the parallel vector potential and the
parallel magnetic field responses are no longer simply flux functions.
Our electromagnetic zonal flow responses provide 15 meaningful tests
of fully electromagnetic gyrokinetic turbulence codes. We focus on the
small skin depth limit in this section and sections \ref{qnandamps}
and \ref{sec:plots}, but section \ref{qnandamps} also gives large skin
depth results.
	
For a pure electrostatic initial condition (${\tilde \Phi ^{(0)}} \ne
0$, $\tilde B_{||}^{(0)} = 0 = \Aptz$), the usual long wavelength
Rosenbluth and Hinton electrostatic zonal flow result is recovered
with only a small $\beta$ correction as seen from
(\ref{eq6p2}). However, the responses of the parallel vector potential
and parallel magnetic fields will have $\cos \theta$ dependence as
well flux surface averaged responses and are given by
(\ref{eq6p3})-(\ref{eq6p6}). Indeed the poloidally varying and flux
surface averaged responses are comparable for $\Bpat$ and $\Apat$.

The shear Alfv\'{e}n field line plucking initial condition ($\Aptz \ne 0$,
$\tilde B_{||}^{(0)}  = 0 = {\tilde \Phi ^{(0)}}$), also
give stronger flux surface averaged $\Apat$ and $\Bpat$ responses as
can be seen from (\ref{eq6p8})-(\ref{eq6p11}). Because the initial
perturbation is electromagnetic, the $\llangle \tilde \Phi \rrangle$
as well as the $\Apat$ responses are free of any $\beta$ multipliers
as can be seen from (\ref{eq6p7})-(\ref{eq6p9}). The compressional
Alfv\'{e}n responses of (\ref{eq6p10}) and (\ref{eq6p11}) contain
$\beta$ multipliers. Only weak poloidal variation is found in this case for $\Apat$. However, the poloidal variation of $\Bpat$ is important with $b\sim \langle \Bpat  \rangle $.

When the initial perturbation is pure field compression ($\tilde
B_{||}^{(0)} \ne 0$, $\Aptz = 0  =  {\tilde \Phi ^{(0)}}$) the
response of $\llangle \tilde \Phi \rrangle$ appears very large since
it is proportional to $1/\beta$ as can be seen from (6.12). In this
case $\llangle \tilde \Phi \rrangle$ is better viewed as an
order unity response to a $\tilde B_{||}^{(0)}  \propto \beta$
initial condition. Then the smallness of $\tilde B_{||}^{(0)}$ cancels
the $1/\beta$ dependence making $\beta \tilde B_{||}^{(0)}$
independent of $\beta$. The $\Bpat$ response is given by
(\ref{eq6p13}) and (\ref{eq6p14}) with poloidal variation again
weak. However, the poloidal variation of $\Apat$ response is
$1/\eps$  stronger than the flux surface averaged
response as seen from (\ref{eq6p15}) and (\ref{eq6p16}).

%_________________________________________________
%-------------------------------------------------blabla
%_________________________________________________

%% ACKNOWLEDGEMENTS
{\bf Acknowledgments}\\ Work supported by U. S. Department of Energy
grants at DE-FG02-91ER-54109 at MIT.  IP is supported by the
Intenational Career Grant (Dnr. 330-2014-631) of Vetenskapsr{\aa}det,
and Marie Sklodowska Curie Actions, Cofund, Project INCA 600398. This
research topic was suggested to the authors by Alex Schekochihin of
Oxford in March of 2013 while we all enjoyed the hospitality and
support of the Wolfgang Pauli Institute in Vienna,
Austria. Fortunately, our friendship has survived in spite of that
seemingly harmless suggestion. Indeed, P.J.C. is indebted to Alex for
accommodations and hospitality at Merton College during subsequent
collaborative visits to the Rudolf Peierls Centre for Theoretical
Physics at Oxford University in 2016 and 2017. In addition, the
authors collaborated during the Gyrokinetic Theory Working Group
Meeting 2016 in Madrid in September.

\appendix
\section{Endless Integrals} 
\label{AppIntegrals}

 For any quantity $X$ we define the flux surface average as
\begin{equation}
\left\langle X\right\rangle = \frac{\oint \frac{d\theta  X  } {\mathbf
  B_0   \cdot   \nabla \theta} }{\oint
        \frac{d\theta    }{\mathbf{B}_0  
        \cdot   \nabla \theta }},
\label{A1}
\end{equation}
and the transit average as
\begin{equation}
\overline X  = \frac{\oint d\tau  X}{ \oint {d\tau   } } = \frac{\oint \frac{d\theta   X{B_0}  }{ v_{||}  \mathbf{B}_0    \cdot   \nabla \theta } }{\oint \frac{d\theta  B_0  }{  v_{||}  \mathbf{B}_0    \cdot   \nabla \theta } },
\label{A2}
\end{equation}
with $d\tau = d\theta {B_0} / ({v_{||}}\mathbf{B}_0 \cdot \nabla \theta)
$. We allow $X$ to depend on $\vpa$ and take the $\theta $ integrations in
both averages (numerators and denominators) to be over a full poloidal
circuit following a charged particle. In this way $\vpa$ and $\theta $
change signs together at a turning point for trapped particles, and
odd functions of $\vpa$, such as $\overline {{v_{||}}} = 0$ and
$\overline {{v_{||}} / {B_0}} = 0$, result in a vanishing transit
average. Using $R / {R_0}(\psi ) \simeq 1 + \eps (\psi )\cos \theta $,
$\eps = \eps (\psi ) = r(\psi ) / {R_0}$, $\underline{B}_0 \equiv
\left\langle {B_0}\right\rangle$, ${B_0} /\underline{B}_0 \simeq {R_0}
/ R \simeq 1 - \eps \cos \theta + \ord({\eps ^2})$, $\lambda = 2\mu
\underline{B}_0 / {v^2}$, and
\begin{equation}
\xi  = \sqrt {1 - \lambda {B_0} /\underline{B}_0},
\label{A3}
\end{equation}
we have for trapped particles that 
\begin{equation}
\overline \xi   = 0,
\label{A4}
\end{equation}
and 
\begin{equation}
\underline{B}_0\overline {\xi     /    {B_0}}  = 0.
\label{A5}
\end{equation}
	 
For the passing particles we see from (\ref{A1}) and (\ref{A2}) that
\begin{equation}
\underline{B}_0\overline {\xi     /    {B_0}}  = \underline{B}_0  /  \left\langle {B_0}  /  \xi \right\rangle
\label{A6}
\end{equation} 
and
\begin{equation}
\overline {\xi W}  = \left\langle {B_0}W\right\rangle /\left\langle {B_0}  /  \xi \right\rangle ,
\label{A7}
\end{equation}
when $W(\psi ,\theta  )$. For trapped particles $\overline {\xi W} =
0$, while $W = 1$ gives the passing result
\begin{equation}
\overline \xi   = \left\langle {B_0}\right\rangle /\left\langle {B_0}  /  \xi \right\rangle  \simeq \frac{ 2\pi }{ \oint d \theta   /  \xi }\left[ 1 + \ord({\eps  ^2})\right].
\label{A8}
\end{equation}
	
We employ a Shafranov shifted circular flux surface model
\citep{RevPlasmaPhysVol2,helanderBook}. We retain the Shafranov shift
${\Delta _S}$ by taking
\begin{equation}
\mathbf{B}_0    \cdot   \nabla \theta   \simeq \frac{\underline{B}_0}{ q{R_0}\left[ 1 + (\eps   - {\Delta '_S})\cos \theta  \right]},
\label{A9}
\end{equation} 
with ${\Delta '_S} = d{\Delta _S}  /  dr\sim \eps $.  Then ${B_0}  /\mathbf{B}_0    \cdot   \nabla \theta   \simeq q{R_0(\psi)}\left[ 1 - {\Delta '_S}\cos \theta   + \ord({\eps  ^2})\right]$.

There are many integrals that need to be performed. The simple ones
involve combinations of powers of $\vpe$ and $\vpa$ multiplied by a
Maxwellian without a transit average. More complicated ones involve
transit and/or flux surface averages, for example, integrals of the
form $\smallint {d^3}v{f_0}{v_{||}}\overline {{v_{||}}}$. The
integrals involving transit average are most conveniently performed
using $v$ and $\lambda$ variables so that
\begin{equation}
{d^3}v   \to \sum\nolimits_\sigma \frac{ {\sigma  } \pi {B_0}{v^3}dvd\lambda }{\underline{B}_0 |  {v_{||}}  | }\to \frac{2\pi {B_0}{v^3}dvd\lambda }{\underline{B}_0  |  {v_{||}}  | }= \frac{2\pi {B_0}{v^2}dvd\lambda }{\underline{B}_0  |  \xi   | } ,
\label{A10}
\end{equation}
with $\sum\nolimits_\sigma  {\sigma  }   |  {v_{||}}  |  \to     2|  {v_{||}}  |$ to account for both directions of ${v_{||}}$ for integrals over even functions of ${v_{||}}$. 
		
We also see from (\ref{A1}) and (\ref{A2}) that
\begin{equation}
 B_0^{ - 1}\smallint {d^3}v{f_0}{v_{||}}\overline X  = \left\langle \smallint {d^3}v\frac{{{f_0}X}}{{\left\langle {B_0}  /  {v_{||}}\right\rangle }}\right\rangle,
\label{A11}
\end{equation}
and
\begin{equation}
 B_0^{ - 1}\smallint {d^3}v{f_0}{v_{||}}\overline X  = \left\langle \smallint {d^3}v{f_0}X\overline {{v_{||}}  / {B_0}} \right\rangle  = \left\langle B_0^{ - 1}\smallint {d^3}v{f_0}{v_{||}}\overline X \right\rangle,
\label{A12}
\end{equation}
where we use (\ref{A6}). The preceding gives, for example,
\begin{equation}
\smallint {d^3}v{f_0}{v_{||}}\overline {{v_{||}}}  = {B_0}\left\langle B_0^{ - 1}\smallint {d^3}v{f_0}{v_{||}}\overline {{v_{||}}} \right\rangle.
\label{A13}
\end{equation}
In the preceding evaluations and hereafter we will often make use of
\begin{equation}
\left\langle \smallint {d^3}v{f_0}\overline X \right\rangle  = \left\langle \smallint {d^3}v{f_0}X\right\rangle
\label{A14}
\end{equation}
and
\begin{equation}
\left\langle \smallint {d^3}v{f_0}Y\overline X \right\rangle  = \left\langle \smallint {d^3}v{f_0}X\overline Y \right\rangle,
\label{A15}
\end{equation}
for arbitrary gyrophase independent functions $X$ and $Y$.
	
\citet{RH} analytically evaluated $\left\langle \smallint {d^3}v{f_0}{v_{||}}({v_{||}}   - \overline {{v_{||}}} )\right\rangle$ to find 
\begin{equation}
(M /  nT)\left\langle \smallint {d^3}v{f_0}{v_{||}}({v_{||}}   - \overline {{v_{||}}} )\right\rangle  = \gamma {\eps  ^{3/2}}   + \ord({\eps  ^2}),
\label{A16}
\end{equation}
where the numerical constant $\gamma\approx 1.64$ comes from $3\int_0^{1 - \eps  } {d\lambda \overline \xi  }  = 2\left[ 1 - \gamma {\eps  ^{3/2}} + \ord({\eps  ^2})\right]$. It is instructive to obtain their result by first using
\begin{equation}
\left\langle \smallint {d^3}v{f_0}{v_{||}}\overline {{v_{||}}} \right\rangle  = 2\pi \left\langle  {B_0}  /\underline{B}_0\right\rangle \int_0^\infty  d v{f_0}{v^4}\int_0^{1 - \eps  } d \lambda \bar \xi  = (3nT  /  2M)\int_0^{1 - \eps  } d \lambda \bar \xi  \left[ 1 + \ord({\eps  ^2})\right],
\label{A17}
\end{equation}
where
\begin{equation}
4\pi M\int_0^\infty  d v{f_0}{v^4}   = 3nT.
\label{A18}
\end{equation} 
Then using (\ref{A8}) for the passing, with $\bar \xi  = 0$ for the trapped, and $\cos \theta   = 1 - 2{\sin ^2}(\theta  /  2)$ gives
\begin{equation}
\xi  \simeq {\left[ 1 - \lambda (1 - \eps  \cos \theta  )\right]^{1/2}} = \sqrt {\left[ 1 - (1 - \eps  )\lambda \right] - 2\eps  \lambda {{\sin }^2}(\theta  /  2)}.
\label{A19}
\end{equation}
We then let $\alpha  = \theta    /  2$ and introduce
\begin{equation}
{k^2}   = 2\eps  \lambda /\left[ 1 - (1 - \eps  )\lambda \right]
\label{A20}
\end{equation} 
with
\begin{equation}
\lambda  = {k^2}  /\left[ (1 - \eps  ){k^2}   + 2\eps  \right],
\label{A21}
\end{equation} 
to obtain the passing result in terms of a complete elliptic integral of the first kind
\begin{equation}
\oint {\frac{{d\theta  }}{{2\pi \xi }}}  = \frac{{\sqrt 2  k}}{{\pi \sqrt {\eps  \lambda } }}\int\limits_0^{\pi /2} {\frac{{d\alpha }}{{\sqrt {1  - {k^2}{{\sin }^2}\alpha } }}}  = \frac{{2\sqrt {(1 - \eps  ){k^2}   + 2\eps  } }}{{\pi \sqrt {2\eps  } }}K(k) \to \left\{ {\begin{array}{*{20}{c}}
  1+k^2/4 &{k \to 0} \\ 
  {\frac{{\sqrt {1  +  \eps  } }}{{\pi \sqrt {2\eps  } }}\ln \left( {\frac{{16}}{{1 - {k^2}}}} \right)}&{k   \to 1} 
\end{array}} \right. .
\label{A22}
\end{equation}
Using $d\lambda  = 4\eps  kdk  /{\left[ (1 - \eps  ){k^2}   + 2\eps  \right]^2} $ then yields
\begin{equation}
\int_0^{1 - \eps  } {d\lambda \overline \xi  }  = 4\eps  \int_0^1 {\frac{{dkk\overline \xi  }}{{{{\left[ (1 - \eps  ){k^2} + 2\eps  \right]}^2}}}}  = 8\pi \eps  \int_0^1 {\frac{{dkk}}{{{{\left[ (1 - \eps  ){k^2} + 2\eps  \right]}^2}\oint d \theta     /  \xi }}} .
\label{A23}
\end{equation}
Inserting $\oint d \theta     /  \xi$ from
(\ref{A22}) gives
\begin{align}
\int_0^{1 - \eps  } {d\lambda \overline \xi  }  = & 2\pi {(2\eps  )^{3/2}}\int_0^1 {\frac{{dkk}}{{{{\left[ (1 - \eps  ){k^2} + 2\eps  \right]}^{5/2}}2K(k)}}} \label{A24} 
	\\ = &  2{(2\eps  )^{3/2}}\int_0^1 {\frac{{dkk}}{{{{\left[ (1 - \eps  ){k^2} + 2\eps  \right]}^{5/2}}}}} \left\{ (1 - \frac{{{k^2}}}{4}) + \left[ \frac{\pi }{{2K(k)}} - (1 - \frac{{{k^2}}}{4})\right]\right\}.
\nonumber
\end{align}
Using
\begin{equation}
\int_0^1 {\frac{{dkk}}{{{{\left[ (1 - \eps  ){k^2} +   2\eps  \right]}^{5/2}}}}}  = \frac{{ - 1}}{{3(1  - \eps  )}}\int_0^1 {dk\frac{d}{{dk}}\frac{1}{{{{\left[ (1  - \eps  ){k^2} +   2\eps  \right]}^{3/2}}}}}  = \frac{1}{{3(1  - \eps  )}}\left(\frac{1}{{{{(2\eps  )}^{3/2}}}} - \frac{1}{{{{(1  + \eps  )}^{3/2}}}}\right),
\label{A24p1}
\end{equation}
\begin{equation}
\int_0^1 {\frac{{dkk}}{{{{\left[ (1 - \eps  ){k^2} + 2\eps  \right]}^{3/2}}}}}  = \frac{{ - 1}}{{1 - \eps  }}\int_0^1 {dk\frac{d}{{dk}}\frac{1}{{{{\left[ (1 - \eps  ){k^2} + 2\eps  \right]}^{1/2}}}}}  = \frac{1}{{1 - \eps  }}\left(\frac{1}{{\sqrt {2\eps  } }} - \frac{1}{{\sqrt {1 + \eps  } }}\right),
\label{A24p2}
\end{equation}
\begin{equation}
\int_0^1 {\frac{{dk{k^3}}}{{{{\left[ (1 - \eps  ){k^2} + 2\eps  \right]}^{5/2}}}}}  = \frac{1}{{(1 - \eps  )}}\left[ \int_0^1 {\frac{{dkk}}{{{{\left[ (1 - \eps  ){k^2} + 2\eps  \right]}^{3/2}}}} - 2\eps  } \int_0^1 {\frac{{dkk}}{{{{\left[ (1 - \eps  ){k^2} + 2\eps  \right]}^{5/2}}}}} \right] \simeq \frac{{2(1 + 2\eps  )}}{{3{{(2\eps  )}^{1/2}}}} - 1,
\label{A24p3}
\end{equation}
and
\begin{equation}
\int_0^1 {\frac{{dkk}}{{{{\left[ (1 - \eps  ){k^2} + 2\eps  \right]}^{5/2}}}}} \left[ \frac{\pi }{{2K(k)}} - \left(1 - \frac{k^2}{4}\right)\right] \simeq \int_0^1 {\frac{{dk}}{{{k^4}}}} \left[ \frac{\pi }{{2K(k)}} - \left(1 - \frac{k^2}{4}\right)\right]  + \ord(\eps  ) \simeq  - 0.10953,
\label{A24p4}
\end{equation}
gives the result needed to recover (\ref{A16})
\begin{equation}
\int_0^{1 - \eps  } {d\lambda \overline \xi  }   =    \frac{2}{3}\left[ 1  - \gamma {\eps  ^{3/2}} + \ord({\eps  ^2})\right].
\label{A25}
\end{equation}
Using (\ref{A13}) and (\ref{A25}) allows us to generalize the
Rosenbluth and Hinton result to find
\begin{equation}
(M /  nT)\smallint {d^3}v{f_0}{v_{||}}({v_{||}}   - \overline {{v_{||}}} ) = 1 - ({B_0}  /\underline{B}_0)\left[ 1 - \gamma {\eps  ^{3/2}}   + \ord({\eps  ^2})\right] = \eps  \cos \theta   + \gamma {\eps  ^{3/2}}   + \ord({\eps  ^2}).
\label{A26}
\end{equation}
This $\theta $ dependence can also be checked by using $v_{||}^2 =
{v^2}(1 - \lambda {B_0} 
/\underline{B}_0  )$ and
\begin{equation}
{v_{||}}\partial {v_{||}}  /  \partial \theta   =  -  ({v^2}\lambda /  2\underline{B}_0)  \partial {B_0}  /  \partial \theta   \simeq  -     \eps  ({v^2}\lambda /  2)\sin \theta  
\label{A26p1}
\end{equation}
to see that   
\begin{equation}
(\partial /\partial \theta  )\left[(\Bzl/\Bz) \smallint {d^3}v{f_0}{v_{||}}({v_{||}}   - \overline {{v_{||}}} ) \right]\simeq \smallint {d^3}v{f_0}{v_{||}}\partial {v_{||}}  /  \partial \theta   \simeq  - \eps  (nT  /  M)\sin \theta  ,
\label{A26p2}
\end{equation}
which when integrated agrees with (\ref{A25}). 
	
We can account for the different behaviour of the trapped (and barely
passing) and passing particles by using $\overline {{v_{||}}} = 0$ and
estimating ${v_{||}}\sim {\eps  ^{1/2}}v$ and ${\int d ^3}v \propto
{\eps  ^{1/2}}$ for the trapped particles, while using
$\overline {{v_{||}}} \simeq {v_{||}}\left[ 1 + \ord(\eps  )\right]$ for the
passing ones. In (\ref{A26}) these estimates give the order $\eps$
poloidal variation as coming from the passing particles while the order
$\eps^{3/2}$ behaviour is due to the trapped.
	
However, we have to be more careful with a related integral since the
Shafranov shift will enter. Using
\begin{equation}
\overline {\xi \cos \theta  }  = 0,
\label{A27}
\end{equation} 
for the trapped, while noting that the passing particle result depends on the
Shafranov shift we find
\begin{equation}
\overline {\xi \cos \theta  }  \simeq \left[ \oint {d\theta  }  (1 - {\Delta '_S}\cos \theta  )  \cos \theta  \right]/(\oint {d\theta    /  \xi }  ) \simeq  - ({\Delta '_S}\overline \xi  /2)\left[ 1 + \ord({\eps  ^2})\right],
\label{A28}
\end{equation}
where we have used (\ref{A8}). Consequently, when integrated over
$\lambda$ we obtain
\begin{align}
(2M  /  3nT) & \smallint {d^3}v{f_0}{v_{||}}\overline {{v_{||}}\cos \theta  }  \simeq ({B_0}/  \underline{B}_0)  \int_0^{1 - \eps  } d \lambda \overline {\xi \cos \theta  }  \simeq  - \pi {\Delta '_S}({B_0}/  \underline{B}_0)\int_0^{1 - \eps  } d \lambda {(\oint {d\theta    /  \xi }  )^{ - 1}}\nonumber
\\\simeq & - ({\Delta '_S}{B_0}/  2\underline{B}_0)\int_0^{1 - \eps  } d \lambda \overline \xi   \simeq  - ({\Delta '_S}/3)(1  - \eps  \cos \theta   - \gamma {\eps  ^{3/2}} + ...) \simeq \ord(\eps  ).
\label{A29}
\end{align}
Notice that (\ref{A29}) is $\ord(\eps)$ since our shifted circle model
requires ${\Delta '_S} = d{\Delta _S}  / 
dr\sim \eps  $. Consequently, we see that $\overline {\xi \cos
  \theta  } \sim \eps $ for the passing particles as expected.
	
A related integral involves
$\underline{B}_0^{ -
  1}\overline {{B_0}\xi }$, with
$\underline{B}_0^{ -
  1}\overline {{B_0}\xi } = 0$ for the trapped, and
\begin{equation}
\underline{B}_0^{ - 1}\overline {{B_0}\xi }  \simeq \overline \xi      \left[ 1 + \ord({\eps  ^2})\right] \simeq \frac{2\pi} {\oint {d\theta    /  \xi } },
\label{A30}
\end{equation}
for the passing particles, where we can use (\ref{A8}) and (\ref{A22}). As a
result, expanding and neglecting ${\eps  ^2}$ corrections yields
\begin{align}
(M/ &   nT{B_0})\smallint {d^3}v{f_0}{v_{||}}( \overline {{B_0}} {v_{||}}   - \overline {{B_0}{v_{||}}} ) \label{A31} 
	\\ \simeq & \gamma {\eps  ^{3/2}} + \eps  \cos \theta   - \eps  (M /nT)\smallint {d^3}v{f_0}(v_{||}^2\overline {\cos \theta  }  - {v_{||}}\overline {{v_{||}}\cos \theta  } )  \simeq \gamma {\eps  ^{3/2}} + \eps  \cos \theta   + \ord({\eps  ^2})\nonumber,
\end{align}
where we use $\overline {\cos \theta  } \sim \eps $ for the
passing and $\overline {\cos \theta  } \sim 1$ for the trapped particles.
	
Another integral of interest is
\begin{equation}
\underline{B}_0^2B_0^{ - 2}\smallint {d^3}v{f_0}v_ \bot ^2{v_{||}}\overline {{v_{||}}}  = 2\pi \int_0^\infty  d v{f_0}{v^6}\int_0^{1 - \eps  } d \lambda \lambda \bar \xi  =  \frac{15n{T^2}  }{  2{M^2}}\int_0^{1 - \eps  } d \lambda \lambda \bar \xi,
\label{A32}
\end{equation}
with $\bar \xi = 0$ for the trapped particles, $\bar \xi$ given by (\ref{A8})
and (\ref{A22}) for the passing particles, and
\begin{equation}
4\pi {M^2}\int_0^\infty  d v{f_0}{v^6}   = 15n{T^2}.
\label{A33}
\end{equation} 
Hence, we need
\begin{equation}
\int_0^{1 - \eps  } {d\lambda \lambda \overline \xi  }  = 2{(2\eps  )^{3/2}}\int_0^1 {\frac{{dk{k^3}}}{{{{\left[ (1 - \eps  ){k^2} + 2\eps  \right]}^{7/2}}}}} \left\{ \left(1 - \frac{{{k^2}}}{4}\right) + \left[ \frac{\pi }{{2K(k)}} - \left(1 - \frac{{{k^2}}}{4}\right)\right]\right\},
\label{A34}
\end{equation} 
along with
\begin{align}
\int_0^1 & {\frac{{dk{k^3}}}{{{{\left[ (1   - \eps  ){k^2} +   2\eps  \right]}^{7/2}}}}}     =    \frac{{ - 1}}{{5(1   - \eps  )}}\int_0^1 {dk{k^2}\frac{d}{{dk}}\frac{1}{{{{\left[ (1   - \eps  ){k^2} +   2\eps  \right]}^{5/2}}}}} \nonumber 
\\ = & \frac{{ - 1}}{{5(1   - \eps  )}}\left[ \frac{1}{{{{(1 + \eps  )}^{5/2}}}} - 2\int_0^1 {\frac{{dkk}}{{{{\left[ (1   - \eps  ){k^2} +   2\eps  \right]}^{5/2}}}}} \right] \simeq \frac{{2(1 + 2\eps   + 3{\eps^2})}}{{15{{(2\eps  )}^{3/2}}}} - \frac{1}{3}, \label{A34p1}
\end{align}
and
\begin{align}
\int_0^1 {\frac{{dk{k^5}}}{{{{\left[ (1  - \eps  ){k^2} +  2\eps  \right]}^{7/2}}}}}  = &\frac{1}{{(1  - \eps  )}}\left[ \int_0^1 {\frac{{dk{k^3}}}{{{{\left[ (1 - \eps  ){k^2} +  2\eps  \right]}^{5/2}}}} - 2\eps  } \int_0^1 {\frac{{dk{k^3}}}{{{{\left[ (1  - \eps  ){k^2} +  2\eps  \right]}^{7/2}}}}}\right] \nonumber
\\ \simeq & \frac{{8(1  + 3\eps  )}}{{15{{(2\eps  )}^{1/2}}}} - 1.
\label{A34p2}
\end{align}
Therefore, we find
\begin{equation}
\int_0^{1 - \eps  } {d\lambda \lambda \overline \xi  }  \simeq \frac{4}{{15}}\left[ 1 - \frac{5}{2}\gamma {\eps  ^{3/2}} + \ord({\eps  ^2})\right]. 
\label{A35}
\end{equation}
As a result,
\begin{equation}
\underline{B}_0^2B_0^{ - 2}\smallint {d^3}v{f_0}v_ \bot ^2{v_{||}}\overline {{v_{||}}}   =    (2n{T^2}  /  {M^2}) \left[ 1 - (5\gamma {\eps  ^{3/2}}/  2) + \ord({\eps  ^2})\right].
\label{A36}
\end{equation}	
	
To perform some of the more complicated integrals we need to
evaluate some transit averages. We have already evaluated $\oint
{d\theta  }   {\xi ^{ - 1}}$ for the passing particles, but now we
also need it for the trapped particles. For the trapped we let
\begin{equation}
\sin (\theta    /  2) = \kappa \sin \alpha,
\label{A37}
\end{equation}
with
\begin{equation}
{\kappa ^2} = {k^{ - 2}}   = \left[ 1 - (1 - \eps  )\lambda \right]/2\eps  \lambda ,
\label{A38}
\end{equation}
and 
\begin{equation}
\lambda  = 1/\left[ (1 - \eps  )   + 2\eps  {\kappa ^2}\right] ,
\label{A39}
\end{equation}
then $\sqrt {{\kappa ^2} - {{\sin }^2}(\theta    /  2)}  = \kappa \cos \alpha $  and $\cos (\theta    /  2)(d\theta    /  2) = \kappa \cos \alpha d\alpha  = \sqrt {1 - {\kappa ^2}{{\sin }^2}\alpha } d\theta    /  2$,
 give the half bounce result
\begin{equation}
\oint {d\theta    /  \xi }  = 4{(2\eps  \lambda )^{ - 1/2}}\int_0^{\pi /2} {d\alpha } /  \sqrt {1 - {\kappa ^2}  {{\sin }^2}\alpha }  = 4{(2\eps  )^{ - 1/2}}\sqrt {(1 - \eps  ) + 2\eps  {\kappa ^2}} K(\kappa ).
\label{A40}
\end{equation}

We also need to evaluate $\overline {\cos \theta }$ for the trapped
particles, but we also evaluate it for the passing particles to check
our estimates.  For the passing we use ${k^2}\cos \theta = {k^2} - 2 +
{\kern 1pt} 2 \left[ 1 - {k^2}{\sin ^2}(\theta / 2)\right]$ to find
\begin{align}
\overline {\cos \theta  }  \simeq & \frac{{\int_{ - \pi }^\pi  {d\theta  \cos \theta    /  \xi } }}{{\int_{ - \pi }^\pi  {d\theta    /  \xi } }} + \ord(\eps  ) \simeq 1 - \frac{2}{{{k^2}}} + \frac{{2\int_0^{\pi /2} d \theta  \sqrt {1 - {k^2}{{\sin }^2}\alpha } }}{{{k^2}\int_0^{\pi /2} {d} \theta  /  \sqrt {1 - {k^2}{{\sin }^2}\alpha } }}\nonumber
	\\ = & 1 - \frac{2}{{{k^2}}} + \frac{{2E(k)}}{{{k^2}K(k)}}   \to \left\{ {\begin{array}{*{20}{c}}
  {7{k^2}/  16}&{k \to 0} \\ 
  { - 1}&{k \to 1} 
\end{array}} \right.\label{A41},
\end{align}
where $E$ is a complete elliptic integral of the second kind. For the
freely passing particles $\lambda \to 0$ giving ${k^2} \to 2\eps
\lambda \to 0$ so we recover the estimate $\overline {\cos \theta }
\sim \eps $. For the trapped particles we again use
(\ref{A37})--(\ref{A39}) to find
\begin{align}
\overline {\cos \theta  }  \simeq & \frac{{\oint {d\theta  \cos \theta  /  \xi } }}{{\oint {d\theta    /  \xi } }} + \ord(\eps  ) \simeq  - 1 + \frac{{2\int_0^{\pi /2} {d\alpha } \sqrt {1 - {\kappa ^2}{{\sin }^2}\alpha } }}{{\int_0^{\pi /2} {d\alpha } /  \sqrt {1 - {\kappa ^2}  {{\sin }^2}\alpha } }} \nonumber
\\ = & \frac{{2E(\kappa )}}{{K(\kappa )}} - 1 \to \left\{ {\begin{array}{*{20}{c}}
  {1 - {\kappa ^2}}&{\kappa  \to 0} \\ 
  { - 1 + \frac{1}{2}\ln (\frac{{16}}{{1 - {\kappa ^2}}})}&{\kappa  \to 1} 
\end{array}} \right.	,
\label{A42}
\end{align}
where we recover the estimate $\overline {\cos \theta } \sim 1$. The
preceding gives the passing particle result
\begin{equation}
\oint d \theta  {\xi ^{ - 1}}\cos \theta   = \frac{{4\sqrt {(1 - \eps  ){k^2}   + 2\eps  } }}{{\sqrt {2\eps  }  {k^2}}}\left[ 2E(k) - (2 - {k^2})K(k)\right],
\label{A43}
\end{equation}
and the half-bounce trapped particle result
\begin{equation}
\oint d \theta  {\xi ^{ - 1}}\cos \theta   = 4{(2\eps  )^{ - 1/2}}\sqrt {(1 - \eps  ) + 2\eps  {\kappa ^2}} \left[ 2E(\kappa ) - K(\kappa )\right],
\label{A44}
\end{equation}	

In addition, we will need $\oint {d\theta   {\xi ^{ - 1}}{{\sin
    }^2}\theta      }$ for the trapped and
passing. Using
\begin{equation}
\sin {  ^2}\theta     = 1 - {\left[ 1 - 2{\sin ^2}(\theta    /  2)\right]^2} = 1 - {\left[ 2(1 - {\kappa ^2}{\sin ^2}\alpha ) - 1\right]^2} =  -  4(1 - {\kappa ^2}{\sin ^2}\alpha ){\kappa ^2}{\sin ^2}\alpha ,
\label{A44p1}
\end{equation}
gives the half-bounce trapped particle result
\begin{align}
{\oint {d\theta  \xi } ^{ - 1}}{\sin ^2}\theta   = & - \frac{{16{\kappa ^2}}}{{\sqrt {2\eps  } }}\sqrt {1 - \eps   + 2\eps  {\kappa ^2}} \int_0^{\pi /2} {d\alpha }   {\sin ^2}\alpha \sqrt {1 - {\kappa ^2}  {{\sin }^2}\alpha } \nonumber
	\\ = &  - \frac{{16}}{{3\sqrt {2\eps  } }}\sqrt {(1 - \eps  ) + 2\eps  {\kappa ^2}} \left[ (1 - {\kappa ^2})K(\kappa ) + (2{\kappa ^2} - 1)E(\kappa )\right],
\label{A45}
\end{align}
where we use \#2.583.4 on p. 182 of \citet{gradshteyn2007}. We will
also need to evaluate $\oint {d\theta   {\xi ^{ - 1}}{{\sin
    }^2}\theta      }$ for the passing (because
of the barely passing particle  contribution). Using
\begin{equation}
\sin {  ^2}\theta     = 1 - {\left[ 1 - 2{\sin ^2}(\theta    /  2)\right]^2} = 4{k^{ - 2}}{\sin ^2}\alpha \left[ (1 - {k^2}{\sin ^2}\alpha ) + {k^2}   - 1\right],
\label{A45p1}
\end{equation}
gives the passing particle result
\begin{align}
\oint & d\theta  \xi^{ - 1} \sin ^2 \theta   =  \frac{16\sqrt {(1 - \eps  )k^2 + 2\eps  } }{k^2\sqrt{2\eps  } }\int_0^{\pi /2} d\alpha \sin^2 \alpha \left[ \sqrt{1 - k^2\sin^2 \alpha }  - \frac{1 - k^2}{\sqrt {1 -k^2 \sin^2\alpha } }\right]\nonumber
\\ = & \frac{16\sqrt {(1 - \eps  )k^2  + 2\eps  } }{k^2 \sqrt {2\eps  } }\int_0^{\pi /2} d\alpha \left[   ( \sin ^2\alpha  + \frac{1 - k^2}{k^2})
\sqrt {1 - k^2 \sin^2\alpha }  - \frac{1 - k^2}{k^2\sqrt {1 - k^2 \sin^2 \alpha } }\right] \nonumber
\\ = & \frac{16 \sqrt {(1 - \eps  )k^2  + 2\eps } }{3k^4\sqrt{2\eps  } } \left[ (2 - k^2)E(k) - 2(1 - k^2)K(k)\right].
\label{A46}
\end{align}	
	
Using the preceding we can evaluate more complicated integrals like
\begin{align}
(  {M^2}\underline{B}_0^2  /  2n{T^2}B_0^2)\smallint {d^3}v{f_0}v_ \bot ^2\overline {v_{||}^2}   = &  ({M^2}{\underline{B}_0}  /  2n{T^2}{B_0})\smallint {d^3}v{f_0}{v^4}\lambda (1 - \lambda  + \eps  \lambda \overline {\cos \theta  } ) \nonumber
	\\ = & 1 - 2\eps  \cos \theta   +    (15\eps      /  4)\int_0^{\underline{B}_0/{B_0}} {d\lambda } {\lambda ^2}{\xi ^{ - 1}}\overline {\cos \theta  },
\label{A46p1}
\end{align}
where we note that
\begin{equation}
\left\langle ({M^2}\underline{B}_0^2  /  2n{T^2}B_0^2)\smallint {d^3}v{f_0}v_ \bot ^2\overline {v_{||}^2} \right\rangle   =  1 +   \ord({\eps  ^2}).
\label{A46p2}
\end{equation}
To evaluate any $\ord({\eps  ^{3/2}})$ terms we Fourier decompose
keeping only the leading harmonic,
\begin{equation}
\varsigma \cos \theta   \simeq (15 /  4\sqrt \eps   )\int_0^{\underline{B}_0/{B_0}} d \lambda {\lambda ^2}{\xi ^{ - 1}}\overline {\cos \theta  } .
\label{A46p3}
\end{equation}
Then we can determine $\varsigma$ from
\begin{equation}
\varsigma  \simeq (15 /  4\pi \sqrt \eps   )\oint d \theta  \cos \theta  \int_0^{\underline{B}_0/{B_0}} d \lambda {\lambda ^2}{\xi ^{ - 1}}\overline {\cos \theta  }  = (15 /  4\pi \sqrt \eps   )\int_0^{1 + \eps  } d \lambda {\lambda ^2}\overline {\cos \theta  } \oint d \theta  {\xi ^{ - 1}}\cos \theta .
\label{A46p4}
\end{equation}
Using the preceding results for $\oint d \theta  {\xi ^{ - 1}}\cos \theta $ and $\overline {\cos \theta  }$ yields
\begin{align}
\varsigma  \simeq &\frac{{30\sqrt 2 }}{\pi }\left\{ \int\limits_0^1 {d\kappa \frac{{\kappa {{\left[ 2E(\kappa ) - K(\kappa )\right]}^2}}}{{{{(1 - \eps   + 2\eps  {\kappa ^2})}^{5/2}}K(\kappa )}}}  + \int\limits_0^1 {dk\frac{{{{\left[ 2E(k) - (2 - {k^2})K(k)\right]}^2}}}{{{{\left[ (1 - \eps  ){k^2} + 2\eps  \right]}^{5/2}}kE(k)}}} \right\} \nonumber
	\\ \simeq & \frac{{30\sqrt 2 }}{\pi }\left\{ \int\limits_0^1 {\frac{{d\kappa \kappa }}{{K(\kappa )}}{{\left[ 2E(\kappa ) - K(\kappa )\right]}^2}}  + \int\limits_0^1 {\frac{{dk}}{{{k^6}E(k)}}{{\left[ 2E(k) - (2 - {k^2})K(k)\right]}^2}} \right\}  \approx 5.294 ,
\label{A47}
\end{align}
where the last integral is well behaved since ${\left[ 2E(k) - (2 -
    {k^2})K(k)\right]^2} \propto {k^8}$ at small $k$, and the $k$
integral is the passing particle contribution and the $\kappa$ integral
the trapped particle contribution.  Consequently,
\begin{equation}
(  {M^2}\underline{B}_0^2  /  2n{T^2}B_0^2)\smallint {d^3}v{f_0}v_ \bot ^2\overline {v_{||}^2}   =  1 - (2\eps   - \varsigma {\eps  ^{3/2}})\cos \theta   + \ord({\eps  ^2}).
\label{A48}
\end{equation}

Combining (\ref{A36}) and (\ref{A46}) gives
\begin{equation}
\underline{B}_0^2B_0^{ - 2}\smallint {d^3}v{f_0}v_ \bot ^2({v_{||}}\overline {{v_{||}}}   - \overline {v_{||}^2} ) =    (2n{T^2}  /  {M^2}) \left[ (2\eps   - \varsigma {\eps  ^{3/2}})\cos \theta   - (5\gamma {\eps  ^{3/2}}/  2) + \ord({\eps  ^2})\right].
\label{A49}
\end{equation}

Another related integral that we will need is
\begin{equation}
\smallint {d^3}v{f_0}v_ \bot ^2\left[  (v_{||}^2 + \overline {v_{||}^2} )       - 2{v_{||}}\overline {{v_{||}}} \right] =  \smallint {d^3}v{f_0}v_ \bot ^2\left[  {({v_{||}} - \overline {{v_{||}}} )^2}   + \overline {{{({v_{||}} - \overline {{v_{||}}} )}^2}} \right] =  \ord({\eps  ^{3/2}}).
\label{A49p1}
\end{equation}
Using $(v_{||}^2 + \overline {v_{||}^2} )  -
2{v_{||}}\overline {{v_{||}}} = 2(\overline {v_{||}^2} -
{v_{||}}\overline {{v_{||}}} ) + (v_{||}^2 - \overline {v_{||}^2} )$
and recalling that to evaluate (\ref{A48}) we used
\begin{equation}
({M^2}\underline{B}_0^2  /2n{T^2}B_0^2)\smallint {d^3}v{f_0}v_ \bot ^2(v_{||}^2 - \overline {v_{||}^2} )     \simeq 4\eps  \cos \theta   +\dots,
\label{A50}
\end{equation} 
then we see that
\begin{equation}
({M^2}\underline{B}_0^2  /  2n{T^2}B_0^2)\smallint {d^3}v{f_0}v_ \bot ^2\left[ {v_{||}}\overline {{v_{||}}}  -  (v_{||}^2 + \overline {v_{||}^2} )  /  2\right] \simeq  - 5\gamma {\eps  ^{3/2}}/  2 -\varsigma \eps^{3/2} \cos \theta+ \dots.
\label{A51}
\end{equation}
	
We also require the $\ord(\eps)$ flux function
\begin{align}
\underline{B}_0^2  & B_0^{ - 2}\smallint {d^3}v{f_0}v_ \bot ^2{v_{||}}\overline {{v_{||}}\cos \theta  }  = (15n{T^2}/  2{M^2})\int_0^{1 - \eps  } d \lambda \lambda \overline {\xi \cos \theta  } \label{A52}
	\\ \simeq & -     (15{\Delta '_S}n{T^2}/  4{M^2})\int_0^{1 - \eps  } d \lambda \lambda \overline \xi   \simeq  -     ({\Delta '_S}n{T^2}/  {M^2})\left[ 1 - (5\gamma {\eps  ^{3/2}}/  2) + ...\right] \simeq  -     ({\Delta '_S}n{T^2}/  {M^2}),
\nonumber
\end{align} 
where we have used (\ref{A18}), (\ref{A28}), and (\ref{A35}).

The more complicated, poloidally dependent integral $\smallint
{d^3}v{f_0}v_ \bot ^2\overline {v_{||}^2\cos \theta  }  
 $ is required to $\ord(\eps^{3/2})$. Flux surface averaging
gives
\begin{align}
\left\langle \underline{B}_0^2 B_0^{ - 2}\smallint {d^3}v{f_0}v_ \bot
^2\overline {v_{||}^2\cos \theta } \right\rangle = &
\underline{B}_0\left\langle \cos \theta \smallint
          {d^3}v{f_0}{v^2}\lambda v_{||}^2 \overline {B_0^{ - 1}}
          \right\rangle \simeq \left\langle \cos \theta \smallint
                        {d^3}v{f_0}{v^2}\lambda v_{||}^2(1 + \eps
                        \overline {\cos \theta } ) \right\rangle
                        \nonumber
\\ \simeq & \underline{B}_0\left\langle B_0^{ - 1}\cos \theta
\right\rangle \smallint {d^3}v{f_0}v_ \bot ^2v_{||}^2 + \eps
\left\langle \cos \theta \smallint {d^3}v{f_0}{v^2}\lambda
v_{||}^2\overline {\cos \theta }  \right\rangle \nonumber
\\ \simeq & (n{T^2}/ {M^2})\left[ 2\eps - {\Delta '_S} + \ord({\eps
    ^2})\right].
\label{A53}
\end{align}
We also need the $\cos \theta $ dependent $\ord(\eps^{3/2})$ portion
of this poloidally dependent integral. Fourier decomposing by keeping
only the fundamental
\begin{equation}
\chi {\eps  ^{3/2}}\cos \theta   \simeq -({M^2}\underline{B}_0^2 /  n{T^2}B_0^2)    \smallint {d^3}v{f_0}v_ \bot ^2\overline {v_{||}^2\cos \theta  }       + \left\langle ({M^2}\underline{B}_0^2 /  n{T^2}B_0^2)    \smallint {d^3}v{f_0}v_ \bot ^2\overline {v_{||}^2\cos \theta  }   \right\rangle,
\label{A53p1}
\end{equation}
we evaluate the Fourier coefficient by multiplying by $q{R_0}\underline{B}_0^{ - 1}\cos \theta  \mathbf{B}_0    \cdot   \nabla \theta $ and flux surface averaging to find
\begin{align}
\chi {\eps  ^{3/2}}/2 \simeq & -(q{R_0}{M^2}/  n{T^2})  \left\langle   B_0^{ - 1}\cos \theta  \mathbf{B}_0    \cdot   \nabla \theta  \smallint {d^3}v{f_0}{v^2}\lambda \overline {v_{||}^2\cos \theta  }   \right\rangle      \simeq \ord({\eps  ^{3/2}})\nonumber
	\\ \simeq & -(q{R_0}{M^2}/  n{T^2})  \left\langle   \cos \theta  \smallint {d^3}v{f_0}{v^2}\lambda v_{||}^2\overline {B_0^{ - 1}\cos \theta  {{\mathbf B}_0}    \cdot   \nabla \theta  }   \right\rangle   \nonumber
	\\ \simeq & -({M^2}/  n{T^2})  \left\langle   \cos \theta  \smallint {d^3}v{f_0}{v^2}\lambda v_{||}^2\overline {\cos \theta  }   \right\rangle  + \ord({\eps  ^2})\nonumber
	\\ \simeq & -(15/  4\pi )\oint {d\theta  } \cos \theta  \int_0^{\underline{B}_0/{B_0}} d \lambda \lambda \xi \overline {\cos \theta  }  \nonumber   
        \\ \simeq &- (15/    4\pi )\int_0^{1 + \eps  } d \lambda \lambda \overline {\cos \theta  }   \oint {d\theta      } \xi \cos \theta      \nonumber
	\\ \simeq & -(15  /  4\pi )\int_0^{1 + \eps  } d \lambda \lambda \overline {\cos \theta  }   \oint {d\theta      } \xi \partial (\sin \theta  )  /  \partial \theta  \nonumber
	\\ \simeq  & +    (15  /  4\pi )\int_0^{1 + \eps  } d \lambda \lambda \overline {\cos \theta  }   \oint {d\theta   \sin \theta      } \partial \xi   /  \partial \theta  \nonumber
	\\ \simeq &   -  (15\eps  /    8\pi )\int_0^{1 + \eps  } d \lambda {\lambda ^2}\overline {\cos \theta  }   \oint {d\theta   {\xi ^{ - 1}}{{\sin }^2}\theta      } .
\label{A53p2}
\end{align}
Inserting $\oint {d\theta   {\xi ^{ - 1}}{{\sin
    }^2}\theta     }$ and $\overline {\cos
  \theta  }$ for the trapped and passing particles gives
\begin{align}
\chi  \simeq & -\frac{{20\sqrt 2 }}{{    \pi }} \left\{ \int_0^1 d \kappa \frac{{\kappa  \left[ 2E(\kappa ) - K(\kappa )\right]}}{{{{(1 - \eps   + 2\eps  {\kappa ^2})}^{7/2}}K(\kappa )}}\left[ (1 - {\kappa ^2})K(\kappa ) + (2{\kappa ^2} - 1)E(\kappa )\right]\right. + \nonumber
\\ \int_0^1 & \left. d k\frac{{ \left[ 2E(k) - (2 - {k^2})K(k)\right]}}{{{{\left[ (1 - \eps  ){k^2} + 2\eps  \right]}^{7/2}}kK(k)}}\left[ (2 - {k^2})E(k) - 2(1 - {k^2})K(k)\right]\right\} \label{A54}
\\ \simeq & -\frac{{20\sqrt 2 }}{{    \pi }}\left\{ \int_0^1 {\frac{{d\kappa \kappa }}{{K(\kappa )}}} \left[ 2E(\kappa ) - K(\kappa )\right]\left[ (1 - {\kappa ^2})K(\kappa ) + (2{\kappa ^2} - 1)E(\kappa )\right] + \right .\nonumber
\\  \int_0^1 & \left. {\frac{{dk}}{{{k^8}K(k)}}} \left[ 2E(k) - (2 - {k^2})K(k)\right]\left[ (2 - {k^2})E(k) - 2(1 - {k^2})K(k)\right]\right\}\approx 0.1131,
\nonumber
\end{align}
where in the last integral $\left[ 2E(k) - (2 -
  {k^2})K(k)\right]\left[ (2 - {k^2})E(k) - 2(1 - {k^2})K(k)\right]
\propto {k^8}$ at small $k$ to keep it well behaved, and the $k$
integral is the passing particle contribution and the $\kappa$
integral is the trapped particle contribution (a notation used from
here on). Then the preceding gives
\begin{equation}
({M^2}\underline{B}_0^2 /  n{T^2}B_0^2)    \smallint {d^3}v{f_0}v_ \bot ^2\overline {v_{||}^2\cos \theta  }     \simeq  (2\eps    -  {\Delta '_S}) - \chi {\eps  ^{3/2}}  \cos \theta   + \ord({\eps  ^2}).
\label{A55}
\end{equation}

We next define the more involved integral I and approximate it by
\begin{equation}
I \equiv ({M^2}/3  n{T^2})\smallint {d^3}v{f_0}{v_{||}}\left[ 3{v_{||}}(\overline {v_{||}^2}  - {v_{||}}\overline {{v_{||}}} ) + (v_{||}^3 - \overline {v_{||}^3} )\right] \simeq {\eps  ^{5/2}}(\sigma  + \upsilon \cos \theta  ),
\label{A56}
\end{equation}
with $\sigma$ and $\upsilon$ order unity constants. The form $I$ is
then rewritten using
\begin{equation}
\int {d^3}v{f_0}v_{||}^2(\overline {v_{||}^2}    -  {v_{||}}{\bar v_{||}})  = \int {d^3}v{f_0}v_{||}^2\left\{  \overline {{{\left[ {{\bar v}_{||}}  + ({v_{||}}   - {{\bar v}_{||}})\right]}^2}}   - {v_{||}}{\bar v_{||}}\right\}   = \int {d^3}v{f_0}v_{||}^2\left[  \overline {{{({v_{||}}   - {{\bar v}_{||}})}^2} }    - {\bar v_{||}}({v_{||}}  - {\bar v_{||}})\right]
\label{A57}
\end{equation}
and
\begin{equation}
\int {d^3}v{f_0}{v_{||}}\frac{v_{||}^3 - \overline {v_{||}^3  }}{3} = \int {d^3}v{f_0}{v_{||}}\left\{ \frac{ {({v_{||}} - {\bar v_{||}})^3}   - \overline {{{({v_{||}} - {{\bar v}_{||}})}^3}} }{ 3} + {v_{||}}{\bar v_{||}}({v_{||}}   - {\bar v_{||}}) - {\bar v_{||}}\overline {{{({v_{||}}   - {{\bar v}_{||}})}^2}} \right\} ,
\label{A58}
\end{equation}
to find that
\begin{equation}
I = ({M^2}/  3n{T^2})\int {d^3}v{f_0}{v_{||}}\left\{ \left[ {({v_{||}} - {\bar v_{||}})^3} - \overline {{{({v_{||}} - {{\bar v}_{||}})}^3}}   \right] + 3  ({v_{||}}   - {\bar v_{||}})\overline {{{({v_{||}}   - {{\bar v}_{||}})}^2}} \right\}  \simeq \ord({\eps  ^{5/2}})
\label{A59}
\end{equation}
and
\begin{align}
\left\langle I\right\rangle  = & ({M^2}/3  n{T^2})\left\langle \smallint {d^3}v{f_0}{v_{||}}\left[ 3{v_{||}}(\overline {v_{||}^2}  - {v_{||}}{\bar v_{||}}) + (v_{||}^3 - \overline {v_{||}^3} )  \right]\right\rangle \nonumber
	\\ = & ({M^2}/  3n{T^2})\left\langle \smallint {d^3}v{f_0}\left[ {({v_{||}}   - {\bar v_{||}})^4} + 3  {({v_{||}}   - {\bar v_{||}})^2}\overline {{{({v_{||}}   - {{\bar v}_{||}})}^2}} \right]\right\rangle \label{A60}
	\\ = & ({M^2}/  3n{T^2})\left\langle \smallint {d^3}v{f_0}\left[ \overline {{{({v_{||}}   - {{\bar v}_{||}})}^4}}  + 3  {({v_{||}}   - {\bar v_{||}})^2}\overline {{{({v_{||}}   - {{\bar v}_{||}})}^2}} \right]\right\rangle  \simeq \sigma {\eps  ^{5/2}},
\nonumber
\end{align}
with the domain of integration of order $\eps^{1/2}$ for the trapped
particles and order unity for the passing particles.  We use
$\overline {{v_{||}}} = 0$ for the trapped particles, estimate
${v_{||}}\sim {\eps ^{1/2}}v$ for the trapped (and barely passing)
particles, and use $\overline {{v_{||}}} \simeq {v_{||}}\left[ 1 +
  \ord(\eps )\right]$ for the freely passing particles. Therefore, we
anticipate that the trapped particle ($\sim\eps^{5/2}$) contributions
will give the final form of (\ref{A55}), with passing particle
contributions $\sim\eps^{3}$.
	
To verify this more completely we next form
\begin{equation}
{B_0}\partial (B_0^{ - 1}I)/  \partial \theta   = ({M^2}{B_0}/  n{T^2}\underline{B}_0)\smallint {d^3}v{f_0}{v_{||}}(\partial {v_{||}}  /  \partial \theta  )\left[ {({v_{||}} - {\bar v_{||}})^2}   + \overline {{{({v_{||}}   - {{\bar v}_{||}})}^2}} \right] \simeq \ord({\eps  ^{5/2}}),
\label{A61}
\end{equation}
where $\theta $ derivatives of the $\lambda$ limits do not
contribute because terms containing transit averages have $\theta $
independent $\lambda$ limits and those not containing transit averages
are multiples of ${v_{||}}$ that will vanish at the upper limit of
$\lambda =
\underline{B}_0  /{B_0}$. Using $v_{||}^2 = {v^2}(1 - \lambda
{B_0} 
/\underline{B}_0  )$ and ${v_{||}}\partial {v_{||}}  /{\kern
  1pt} \partial \theta  = -  ({v^2}\lambda / 
2\underline{B}_0)  \partial {B_0}  /  \partial
\theta  \simeq -     \eps  ({v^2}\lambda
/  2)\sin \theta $ gives
\begin{align}
{B_0} & \partial (B_0^{ - 1}I)/  \partial \theta   \simeq  - \eps  ({M^2}{B_0}/  2n{T^2}\underline{B}_0)\sin \theta  \smallint {d^3}v{f_0}\lambda {v^2}\left[ {({v_{||}} - {\bar v_{||}})^2}   + \overline {{{({v_{||}}   - {{\bar v}_{||}})}^2}} \right] \simeq \ord({\eps  ^{5/2}})\nonumber
	\\ \simeq &  - \eps  ({M^2}/  2n{T^2})\sin \theta  \smallint {d^3}v{f_0}v_ \bot ^2(v_{||}^2   - 2{v_{||}}{\bar v_{||}}   + \overline {v_{||}^2} ) \simeq  - (5\gamma  + \varsigma \cos   \theta  ){\eps  ^{5/2}}\sin \theta  .
\label{A62}
\end{align}
where we have used (\ref{A36}) and (\ref{A48}). Integrating gives to
lowest order
\begin{equation}
I \simeq {\eps  ^{5/2}}\left[ \sigma   + 5\gamma \cos \theta   + (\varsigma     /  4)\cos 2\theta  \right],
\label{A63}
\end{equation}
giving $\upsilon = 5\gamma$ and $\sigma$ given by (\ref{A60}) to be
the constant found by evaluating
\begin{align}
\sigma  \simeq & (5  /  2{\eps  ^{5/2}})\left\langle \int_0^{\underline{B}_0/{B_0}} d \lambda {\xi ^{ - 1}}\left[ {(\xi    - \bar \xi )^4} + 3{(\xi    - \bar \xi )^2}\overline {{{(\xi    - \bar \xi )}^2}} \right]\right\rangle \nonumber
	\\ \simeq & (5  /  2{\eps  ^{5/2}})\left\langle \int_0^{\underline{B}_0/{B_0}} d \lambda {\xi ^{ - 1}}\left[ \overline {{{(\xi    - \bar \xi )}^4}}  + 3{(\xi    - \bar \xi )^2}\overline {{{(\xi    - \bar \xi )}^2}} \right]\right\rangle \label{A64}
	\\ \simeq & (5  /  2{\eps  ^{5/2}})\left\{ \int_0^{1 + \eps  } d \lambda \overline {{{(\xi    - \bar \xi )}^4}} \oint {\frac{{d\theta  }}{{2\pi \xi }}}  + 3\int_0^{1 + \eps  } d \lambda \overline {{{(\xi    - \bar \xi )}^2}} \left[ \oint {\frac{{d\theta  \xi }}{{2\pi }}}    - 2\bar \xi  + {\overline \xi  ^2}\oint {\frac{{d\theta  }}{{2\pi \xi }}} \right]\right\} .
\nonumber
\end{align}
The $\varsigma     \cos 2\theta $ term in
(\ref{A63}) is of no consequent for our purposes since we only keep
$\cos \theta $ terms.
	
We also need 
\begin{align}
{I_h} \equiv ({M^2}/3  n{T^2})\smallint {d^3}v{f_0}{v_{||}}\left[ 3{v_{||}}h(\overline {v_{||}^2{h^2}}  - {v_{||}}h\overline {{v_{||}}h} ) + (v_{||}^3{h^3} - \overline {v_{||}^3{h^3}} )\right] \label{A65}
          \\ = ({M^2}/  3n{T^2})\smallint {d^3}v{f_0}{v_{||}}\left\{ \left[ {({v_{||}}h - \overline {{v_{||}}h} )^3} - \overline {{{({v_{||}}h - \overline {{v_{||}}h} )}^3}}   \right] + 3  ({v_{||}}h   - \overline {{v_{||}}h} )\overline {{{({v_{||}}  h - \overline {{v_{||}}h} )}^2}} \right\}  \simeq \ord({\eps  ^{5/2}}),
\nonumber
\end{align} 
where we define $h = \underline{B}_0 / {B_0} \simeq 1 + \eps \cos
\theta $ and note that for the trapped particles $\overline
       {{v_{||}}h} = 0$. Using ${v_{||}}h - \overline {{v_{||}}h}
       \simeq {v_{||}} - \overline {{v_{||}}} + \eps {v_{||}}\cos
       \theta - \eps \overline {{v_{||}}\cos \theta } \simeq {v_{||}}
       - \overline {{v_{||}}} + \eps {v_{||}}\cos \theta + \ord({\eps
         ^2})$ we expand for $\eps\ll 1$ to find
\begin{equation}
{I_h} - I \simeq \eps  ({M^2}/  n{T^2})\smallint
{d^3}v{f_0}{v_{||}}\left[ {v_{||}}{({v_{||}} - \overline {{v_{||}}} )^2}\cos
  \theta  +   {v_{||}}\overline {{{({v_{||}}  -
        \overline {{v_{||}}} )}^2}} \cos \theta  \right] \simeq
\ord({\eps  ^3}),
\label{A66}
\end{equation}
where we use ${v_{||}} - \overline {{v_{||}}} \sim v\eps $ for the
passing and ${v_{||}}\sim v{\eps  ^{1/2}}$ for the trapped particles.

Finally, we can simplify (\ref{A64}) further by first forming $\oint
{d\theta  \xi }$. For the passing particles we let $\alpha = \theta 
 /  2$ to find
\begin{equation}
\oint {d\theta  \xi }  = \frac{{4\sqrt {2\eps  } }}{{\sqrt {(1 - \eps  ){k^2}   + 2\eps  } }}\int_0^{\pi /2} d \alpha \sqrt {1 - {k^2}{{\sin }^2}\alpha }  = \frac{{4\sqrt {2\eps  } E(k)}}{{\sqrt {(1 - \eps  ){k^2}   + 2\eps  } }},
\label{A67}
\end{equation}
while for the trapped we let $\sin (\theta    / 
2) = \kappa \sin \alpha$ to obtain the half bounce result
\begin{equation}
\oint {d\theta  \xi }  = \frac{{4\sqrt {2\eps  } {\kappa ^2}}}{{\sqrt {1 - \eps   + 2\eps  {\kappa ^2}} }}\int_0^{\pi /2} {\frac{{d\alpha {{\cos }^2}\alpha }}{{\sqrt {1 - {\kappa ^2}{{\sin }^2}\alpha } }}}  = \frac{{4\sqrt {2\eps  } \left[ E(\kappa )  - (1  - {\kappa ^2})K(\kappa )\right]}}{{\sqrt {1 - \eps   + 2\eps  {\kappa ^2}} }},
\label{A68}
\end{equation}
where we use
\begin{equation}
\int_0^{\pi /2} {\frac{{d\alpha {{\cos }^2}\alpha }}{{\sqrt {1  - {\kappa ^2}{{\sin }^2}\alpha } }}}  = \frac{1}{{{\kappa ^2}}}\left[ E(\kappa ) - (1 - {\kappa ^2})K(\kappa )\right]
\label{A69}
\end{equation}
from no.~2.584.6 on p. 186 of \citet{gradshteyn2007} or use our previous results.
	
Inserting (\ref{A22}), (\ref{A40}), (\ref{A67}) and (\ref{A68}) into
(\ref{A64}) and using $\bar \xi = 0$ for the trapped and (\ref{A8})
for the passing particles gives
\begin{align}
\sigma  \simeq & \frac{{10\sqrt 2 }}{{\pi {\eps  ^2}}}  \left\{ \int_0^1 {\frac{{dkk\overline {{{(\xi    - \bar \xi )}^4}} K(k)}}{{{{\left[ (1 - \eps  ){k^2} + 2\eps  \right]}^{3/2}}}}}  + \int_0^1 {\frac{{d\kappa \kappa \overline {\xi {  ^4}} K(\kappa )}}{{{{\left[ 1 - \eps   + 2\eps  {\kappa ^2}\right]}^{3/2}}}}} \right. \nonumber
\\ + & \left. 6\eps  \int_0^1 {\frac{{dkk\overline {{{(\xi    - \bar \xi )}^2}} }}{{{{\left[ (1 - \eps  ){k^2} + 2\eps  \right]}^{5/2}}}}} \left[   E(k) - \frac{{{\pi ^2}}}{{4K(k)}}\right] + 6\eps  \int_0^1 {\frac{{d\kappa \kappa \overline {\xi {  ^2}} }}{{{{\left[ 1 - \eps    + 2\eps  {\kappa ^2}\right]}^{5/2}}}}} \left[   E(\kappa ) - (1 - {\kappa ^2})K(\kappa )\right]\right\} .
\label{A69p1}
\end{align}
Simplifying by noting that $\overline {{{(\xi   - \bar \xi
      )}^2}} \propto {k^2}$,
\begin{align}
\sigma  \simeq & \frac{{10\sqrt 2 }}{{\pi {\eps  ^2}}}  \left\{ \int_0^1 d k{k^{ - 2}}\overline {{{(\xi    - \bar \xi )}^4}} K(k) + \int_0^1 d \kappa \kappa \overline {\xi {  ^4}} K(\kappa ) \right. \label{A70}
	\\ + & \left. 6\eps  \int_0^1 d k{k^{ - 4}}(\overline {{\xi ^2}}  - {\overline \xi  ^2})\left[   E(k) - \frac{{{\pi ^2}}}{{4K(k)}}\right] + 6\eps  \int_0^1 d \kappa \kappa \overline {{\xi ^2}} \left[   E(\kappa ) - (1 - {\kappa ^2})K(\kappa )\right]\right\} .
\nonumber
\end{align}
	
We can simplify the trapped particle contributions further by using
no.~2.584.15 on p. 187 of \citet{gradshteyn2007},
\begin{equation}
\int_0^{\pi /2} {\frac{{d\alpha {{\cos }^4}\alpha }}{{\sqrt {1  -
        {\kappa ^2}{{\sin }^2}\alpha } }}} = \frac{{4{\kappa ^2} -
    2}}{{3{\kappa ^4}}}E(\kappa ) + \frac{{3{\kappa ^4} - 5{\kappa ^2}
    + 2}}{{3{\kappa ^4}}}K(\kappa ),
\label{A71}
\end{equation}
 to write
\begin{equation}
\overline {\xi {  ^4}}  \simeq \frac{{{{\oint {d\theta  \xi } }^3}}}{{{{\oint {d\theta  \xi } }^{ - 1}}}} \simeq \frac{{4{\eps  ^2}{\kappa ^4}\int_0^{\pi /2} {\frac{{d\alpha {{\cos }^4}\alpha }}{{\sqrt {1  - {\kappa ^2}{{\sin }^2}\alpha } }}} }}{{\int_0^{\pi /2} d \alpha /\sqrt {1  - {\kappa ^2}{{\sin }^2}\alpha } }} \simeq \frac{{4{\eps  ^2}}}{3}\left[ (4{\kappa ^2} - 2)\frac{{E(\kappa )}}{{K(\kappa )}} + (3{\kappa ^4} - 5{\kappa ^2} + 2)\right].
\label{A72}
\end{equation}
Then the first trapped particle term becomes
\begin{equation}
\int_0^1 d \kappa \kappa \overline {\xi {  ^4}} K(\kappa ) \simeq \frac{{4{\eps  ^2}}}{3}\int_0^1 d \kappa \kappa \left[ (4{\kappa ^2} - 2)E(\kappa ) + (3{\kappa ^4} - 5{\kappa ^2} + 2)K(\kappa )\right] = \frac{{32{\eps  ^2}}}{{75}},
\label{A73}
\end{equation}
upon using $\int_0^1 d \kappa \kappa E(\kappa ) = \frac{2}{3}$,
$\int_0^1 d \kappa {\kappa ^3}E(\kappa ) = \frac{{14}}{{45}}$,
$\int_0^1 d \kappa \kappa K(\kappa ) = 1$, $\int_0^1 d \kappa {\kappa
  ^3}K(\kappa ) = \frac{5}{9}$, and $\int_0^1 d \kappa {\kappa
  ^5}K(\kappa ) = \frac{{89}}{{225}}$, from pp. 615-616, no.~5.112.3-7
of \citet{gradshteyn2007}. To simplify the second trapped particle
term we use
\begin{equation}
\overline {\xi {  ^2}}  \simeq \frac{{\oint {d\theta  \xi } }}{{{{\oint {d\theta  \xi } }^{ - 1}}}} \simeq 2\eps  \left[ \frac{{E(\kappa )}}{{K(\kappa )}} - (1 - {\kappa ^2})\right],
\label{A74}
\end{equation}
to find
\begin{align}
\int_0^1 & d \kappa \kappa \overline {{\xi ^2}} \left[   E(\kappa ) - (1 - {\kappa ^2})K(\kappa )\right] \simeq 2\eps  \int_0^1 d \kappa \kappa {\left[   E(\kappa ) - (1 - {\kappa ^2})K(\kappa )\right]^2}/ K(\kappa ),\nonumber
	\\ \simeq &  2\eps  \left\{ \left[ \int_0^1 d \kappa \kappa   \frac{{{E^2}(\kappa )}}{{K(\kappa )}}\right]  - \frac{{32}}{{75}}\right\} ,
\label{A75}
\end{align}
where we use the preceding results to perform some of the integrals. 
	
We can also simplify the simpler of the passing particle terms using
\begin{equation}
\overline {\xi {  ^2}}  \simeq \frac{{\oint {d\theta  \xi } }}{{{{\oint {d\theta  \xi } }^{ - 1}}}} \simeq \frac{{2\eps  E(k)}}{{\left[ (1 - \eps  ){k^2} + 2\eps  \right]K(k)}} \simeq 2\eps  \frac{{E(k)}}{{k^2 K(k)}}
\label{A76}
\end{equation}
and
\begin{equation}
{\overline \xi  ^2} \simeq \frac{{4{\pi ^2}}}{{{{({{\oint {d\theta  \xi } }^{ - 1}})}^2}}} \simeq \frac{{{\pi ^2}\eps  }}{{2\left[ (1 - \eps  ){k^2} + 2\eps  \right]{K^2}(k)}} \simeq \frac{{{\pi ^2}\eps  }}{{2{k^2K^2}(k)}},
\label{A77}
\end{equation}
giving
\begin{equation}
\int_0^1 d k{k^{ - 4}}(\overline {{\xi ^2}}  - {\overline \xi  ^2})\left[   E(k) - \frac{{{\pi ^2}}}{{4K(k)}}\right] \simeq 2\eps  \int_0^1 {\frac{{dk}}{{{k^6}K(k)}}} {\left[   E(k) - \frac{{{\pi ^2}}}{{4K(k)}}\right]^2}.
\label{A78}
\end{equation}

The remaining passing particle term can be rewritten as
\begin{align}
\int_0^1 & d k{k^{ - 2}}\overline {{{(\xi    - \bar \xi )}^4}} K(k) = \int_0^1 d k{k^{ - 2}}(\overline {\xi {  ^4}}  - 4\overline {\xi {  ^3}} \overline {\xi   }  + 6\overline {\xi {  ^2}} {\overline {\xi   } ^2} - 3{\overline {\xi   } ^4})K(k) \nonumber
	\\ = & \int_0^1 d k{k^{ - 2}}\left[ (\overline {\xi {  ^4}}  - 4\overline {\xi {  ^3}} \overline {\xi   }  + 3\overline {\xi {  ^2}} {\overline {\xi   } ^2})   + 3{\overline {\xi   } ^2}(\overline {\xi {  ^2}  }  - {\overline {\xi   } ^2})\right]K(k) \nonumber
	\\ = & 4{\eps  ^2}\int_0^1 {\frac{{dk}}{{{k^6}}}} \left\{ \frac{{2(2 - {k^2})E(k) - (1 - {k^2})K(k)}}{3} + \frac{{{\pi ^2}}}{{4{K^2}(k)}}\left[ 3E(k) - 2(2 - {k^2})K(k)\right]\right\} \label{A79}
	\\ + & 3{\pi ^2}{\eps  ^2}\int_0^1 {\frac{{dk}}{{{k^6}K(k)}}} \left[ \frac{{E(k)}}{{K(k)}} - \frac{{{\pi ^2}}}{{4{K^2}(k)}}\right] \nonumber
	\\ \simeq & 3{\pi ^2}{\eps  ^2}\int_0^1 {\frac{{dk}}{{{k^6}K(k)}}} \left[ \frac{{E(k)}}{{K(k)}} - \frac{{{\pi ^2}}}{{4{K^2}(k)}}\right] + 3{\pi ^2}{\eps  ^2}\int_0^1 {\frac{{dk}}{{{k^6}K(k)}}} \left[ \frac{{E(k)}}{{K(k)}} - 1 \right] \nonumber
	\\ + & \frac{{4{\eps  ^2}}}{3}\int_0^1 {\frac{{dkK(k)}}{{{k^6}}}} \left[ 2(2 - {k^2})\frac{{E(k)}}{{K(k)}} - (1 - {k^2}) - \frac{{3{\pi ^2}}}{{4{K^2}(k)}}(1 - 2{k^2})\right],
\nonumber
\end{align}
by using the passing particle results
\begin{equation}
\bar \xi  \simeq \frac{{2\pi }}{{{{\oint {d\theta  \xi } }^{ - 1}}}} \simeq \frac{{\pi \sqrt {2\eps  } }}{{2\sqrt {(1 - \eps  ){k^2} + 2\eps  } K(k)}} \simeq \frac{{\pi \sqrt {2\eps  } }}{{2kK(k)}},
\label{A80}
\end{equation}
\begin{equation}
\overline {\xi {  ^3}}  \simeq \frac{{\oint {d\theta  {\xi ^2}} }}{{{{\oint {d\theta  \xi } }^{ - 1}}}} \simeq \frac{{\pi \eps  \sqrt {2\eps  } (2 - {k^2})}}{{2{{\left[ (1 - \eps  ){k^2} + 2\eps  \right]}^{3/2}}K(k)}} \simeq \frac{{\pi \eps  \sqrt {2\eps  } (2 - {k^2})}}{{2k^3K(k)}},
\label{A81}
\end{equation}
and
\begin{align}
\overline {\xi {  ^4}}  \simeq & \frac{{\oint {d\theta  {\xi ^3}} }}{{{{\oint {d\theta \xi } }^{ - 1}}}} \simeq \frac{{4{\eps  ^2}}}{{{{\left[ (1 - \eps  ){k^2}   + 2\eps  \right]}^2}K(k)}}\int\limits_0^{\pi /2} {\frac{{d\alpha \left[ {{(1 - {k^2})}^2}  +  2{k^2}(1 - {k^2}) {{\cos }^2}\alpha  + {k^4}{{\cos }^4}\alpha \right]}}{{\sqrt {1 - {k^2}si{n^2}\alpha } }}} \nonumber
       \\ \simeq & \frac{4{\eps  ^2}}{k^4}\left\{ {(1 - {k^2})^2} + \frac{{2(1 - {k^2})}}{{K(k)}}\left[ E(k)  - (1  - {k^2})K(k)\right] + \frac{{\left[ (4{k^2}   - 2)E(k) + (3{k^4}   - 5{k^2} + 2)K(k)\right]}}{{3K(k)}}\right\} \nonumber
	 \\ \simeq & \frac{{4{\eps  ^2}}}{3k^2}\left[ 2(2 - {k^2})\frac{{E(k)}}{{K(k)}} - (1 - {k^2})\right].
\label{A82}
\end{align}

Putting all this together (\ref{A70}) becomes 
\begin{align}
\sigma  = &\frac{40\sqrt 2 }{\pi } \left\{ {3\left[\int_0^1 d \kappa \kappa 
\frac{ E^2(\kappa )}{K(\kappa )}\right] - \frac{88}{75} + 3\int_0^1 
\frac{dk}{k^6 K(k)} \left[E(k) - \frac{\pi^2}{4K(k)}\right]^2} \right.
\label{A83}
\\+ &  \left. \int_0^1 \frac{dk}{k^6} \left[ \frac{2(2 - k^2)E(k)}{3} - 
\frac{(1 -k^2)K(k)}{3} + \frac{3\pi^2}{4K(k)}\left( \frac{2E(k)}{K(k)} 
- \frac{\pi^2}{{4K^2(k)}} -  \frac{2(2 - k^2)}{3}\right)\right] 
\vphantom{\left[E(k) - \frac{\pi^2}{4K(k)}\right]^2}
 \right\}.
 \nonumber
\end{align}
Numerically evaluating the integrals yields the constant $\sigma$ to be
\begin{equation}
\sigma\approx 5.325
\label{A84}
\end{equation}

\bibliographystyle{jpp}

\providecommand{\noopsort}[1]{}\providecommand{\singleletter}[1]{#1}

\end{document}